\documentclass[fleqn,usenatbib]{mnras}

\usepackage{newtxtext,newtxmath}

\usepackage[T1]{fontenc}

\usepackage{subcaption}
\usepackage[dvipsnames]{xcolor}

\DeclareRobustCommand{\VAN}[3]{#2}
\let\VANthebibliography\thebibliography
\def\thebibliography{\DeclareRobustCommand{\VAN}[3]{##3}\VANthebibliography}

\newcommand{\vect}[1]{\boldsymbol{\mathbf{#1}}}

\usepackage{graphicx}	
\usepackage{amsmath}	

\title{A fast neural emulator for interstellar chemistry}

\author[A. Asensio Ramos et al.]{
A. Asensio Ramos$^{1,2}$,\thanks{E-mail: andres.asensio@iac.es}
C. Westendorp Plaza$^{1,2}$,
D. Navarro-Almaida$^{3}$,
P. Rivi\`ere-Marichalar$^{4}$,
\newauthor
V. Wakelam$^{5}$,
A. Fuente$^{6}$
\\
$^{1}$Instituto de Astrof\'isica de Canarias (IAC), Avda V\'ia L\'actea S/N, 38200 La Laguna, Tenerife, Spain\\
$^{2}$Departamento de Astrof\'isica, Universidad de La Laguna, 38205 La Laguna, Tenerife, Spain\\
$^{3}$Universit\'{e} Paris-Saclay, Universit\'e Paris Cit\'{e}, CEA, CNRS, AIM, F-91191 Gif-sur-Yvette, France\\
$^{4}$Observatorio Astron\'omico Nacional (OAN,IGN), Calle Alfonso XII, 3. 28014 Madrid, Spain\\
$^{5}$Laboratoire d'astrophysique de Bordeaux (LAB), Univ. Bordeaux, CNRS, B18N, all\'ee Geoffroy Saint-Hilaire, 33615 Pessac, France\\
$^{6}$Centro de Astrobiolog\'{\i}a (CSIC/INTA), Ctra. de Torrej\'on a Ajalvir km 4, 28850 Torrej\'on de Ardoz, Spain
}

\date{Accepted XXX. Received YYY; in original form ZZZ}

\pubyear{2024}

\begin{document}
\label{firstpage}
\pagerange{\pageref{firstpage}--\pageref{lastpage}}
\maketitle

\begin{abstract}
Astrochemical models are important tools to interpret observations of molecular and atomic species 
in different environments. However, these models are time-consuming, precluding a thorough exploration 
of the parameter space, leading to uncertainties and biased results. Using neural networks to simulate 
the behavior of astrochemical models is a way to circumvent this problem, providing fast calculations 
that are based on real astrochemical models. In this paper, we present a fast neural emulator of the 
astrochemical code Nautilus based on conditional neural fields. The resulting model produces the 
abundance of 192 species for arbitrary times between 1 and 10$^7$ years. Uncertainties well below 
0.2 dex are found for all species, while the computing time is of the order of 10$^4$ smaller
than Nautilus. This will open up the possibility of performing much more complex forward
models to better understand the physical properties of the interstellar medium. As an example of the power 
of these models, we ran a feature importance analysis on the electron abundance predicted by Nautilus. 
We found that the electron density is coupled to the initial sulphur abundance in
a low density gas. Increasing the initial sulphur abundance from a depleted scenario to the cosmic 
abundance leads to an enhancement of an order of magnitude of the electron density. This enhancement 
can potentially influence the dynamics of the gas in star formation sites.
\end{abstract}




\section{Introduction}
Interstellar matter is continuously processed describing a life-cycle that
connects stellar and interstellar environments. Interstellar matter forms  
filamentary clouds which fragment and collapse to form dense cores, where protostars are born.
These protostars develop accretion disks, these latter being at
the origin of planetary systems. The life-cycle is closed when, at the end of their lives, stars
inject a good fraction of their mass back to the interstellar medium, either through gentle
winds, in the case of low-mass stars, or as supernovae explosions, in the case of massive
stars.

Molecules serve as excellent tracers of the physical and chemical evolution of interstellar matter \citep{Jorgensen2020}. To date, more 
than 200 different molecules have been identified in space\footnote{See the Cologne Database for Molecular 
Spectroscopy \citep[CDMS;][]{Muller2001} found at \texttt{https://cdms.astro.uni-koeln.de/classic/molecules}.}.  The molecules observed in space go from the simplest 
hydrides, such as OH, OH$^+$, CH, and CH$^+$, present
in diffuse interstellar clouds \citep{Gerin2016, Gerin2019}, to the complex organic molecules, some of them with prebiotic
relevance, that are detected in pre-stellar cores, young protostars, and planet-forming disks \citep{Guelin2022, Crockett2015, Riviere2020, Oberg2023}. 
In environments that are heavily obscured at ultraviolet, optical,
and near-infrared wavelengths, the observation of molecules becomes the only means to get
access to the physical and chemical processes at work.
Observations have  demonstrated that the chemical nature of molecules varies drastically as objects evolve and 
their environment change. Some molecules are routinely used as chemical diagnostics of
critical aspects such as, e.g., the gas ionization degree (see, e..g., \citealp{Caselli1998, Bergin1999, Fuente2019, Pineda2024}), the local flux of cosmic rays \citep{Favre2017, Cabedo2023}, the gas kinetic
temperature \citep{Hacar2020}, and the density \citep{Navarro-Almaida2021}. However, the reliability of  
molecular diagnostics used to determine the properties of the interstellar medium depends on our ability to 
understand the interstellar chemistry and the predictive power of our chemical models.
Complex chemical models including thousands of reactions are
routinely used in astrophysics \citep{Petit2006, Taquet2012, Ruaud2016, Holdship2017, Laas2019}. 
These codes are time-consuming and 
computationally expensive, which limits the total number of computed models. This limitation challenges the
interpretation of the data provided by current telescopes
which allows to simultaneously observe tens, even hundreds, of molecular lines
\citep{Fuente2014, Fuente2019, McGuire2020, Cernicharo2023}, and map large areas 
in affordable amounts of time \citep{Friesen2017, Gratier2021}.

Molecules have an important
influence on the heating and cooling of interstellar gas and, therefore, on the cloud structure
of the interstellar medium in galaxies. Molecular clouds are the site of ongoing star formation
to a large extent because molecules are efficient gas coolers, thereby diminishing thermal
support relative to self-gravity. Moreover, molecules suppress the degree of ionization in
these clouds and, therefore, decouple the gas from any supporting magnetic fields.
Molecules, thus, play a key role in the formation process of stars and planets.
Steady state models have been widely used
to interpret molecular observations in the last decades. Many of the processes linked to the formation of
a star, such as bipolar outflows and disk formation, have timescales (10$^3$$-$10$^5$ yr) smaller than
the time required for most species to reach a steady state in their abundances ($\sim$1 Myr).
Coupling dynamics and chemistry is therefore of paramount importance to progress in our
understanding of the physical and chemical evolution from the initial cold pre-stellar core to a
protoplanetary disk, where planets are formed.
 
Numerical simulations have become a valuable tool to study, in a realistic manner, the evolution of interstellar matter during the star formation process. This dynamical evolution has been investigated following different approaches: analytical one-dimensional models \citep[see, e.g.,][]{Priestley2018}, two-dimensional semi-analytical models \citep{Drozdovskaya2014, Drozdovskaya2016}, and three-dimensional radiation-magnetohydrodynamical (R-MHD) simulations \citep[see, e.g.,][]{Commercon2012, Masson2016, Hincelin2016,  Zhao2018a, Hennebelle2020, Gong2020, Gong2021, Lebreuilly2023}. MHD simulations have become the state-of-the-art approach for the analysis of the gravitational collapse of prestellar cores into protostellar objects probing the decisive role of the magnetic field in the process of collapse, fragmentation, and disk formation. 
A few of these state-of-the-art simulations already incorporate a reduced chemical network that provides important parameters for MHD calculations, such as the gas ionization degree
\citep{Marchand2016, Zhao2018b}. However, full coupling of the MHD evolution of the gas with a full chemical reaction network with thousands of reactions,
is challenging with the current computational capabilities \citep{Hsu2021, Hsu2023}. 
As a feasible approximation, an alternative
method based on the usage of virtual particles has been used. These particles store the key parameters
required for the computation of the chemistry (namely temperature and density) and a full state-of-the-art
chemical model is then used to predict their chemical composition as a function of time 
\citep{Hincelin2016, Coutens2020, NavarroAlmaida2024}. Yet, these
calculations are expensive in terms of computing time since a large number of particles is needed to sample
all the protostellar components and only a handful of cases have been completed, thus far. Consequently, a
full comparison of these 3D chemo-MHD simulations with observations has not been done yet, so that
their predictive power remains largely unexplored.
 
Astrochemistry has become a necessary tool for understanding the interstellar medium. 
Chemical models consist of a system of coupled ordinary differential equations (ODE)
that represent a chemical network. This system is resolved for a set of input parameters which include 
the gas physical conditions (density, temperature), cosmic ray ionization rate per H atom, incident UV 
field, visual extinction, grain properties, initial chemical composition, and elemental abundances. 
Computing large grids of complex chemical models, varying input parameters to describe all astrophysical 
environments, and predicting the chemistry for each time step is unaffordable. It is therefore interesting 
to develop fast chemistry emulators which allow to address some problems in a fast way, which is precisely the
purpose of this work. We are not the first team to explore emulators in chemistry.
The vast majority of accurate enough emulators are based on neural networks, although other methods
like random forest regression have been applied \citep{2021A&A...645A..28B}.
\cite{2019A&A...630A.117D} used a neural network to emulate astrochemical models in equilibrium after 10$^7$
years obtained with UCLCHEM \citep{2017AJ....154...38H}. \cite{2021A&A...653A..76H} trained
a neural network to predict the change in abundance during a small time step, allowing the
user to evolve the chemistry for long times by iteratively applying the neural network.
\cite{2022A&A...668A.139G} and \cite{2023arXiv231206015S} used autoencoders to
project large chemical networks into a low-dimensional latent space, where the chemistry
is potentially evolved faster. A similar approach is followed by \cite{maes24} for emulating the
chemistry in dynamical environments.
\cite{2023MNRAS.518.5718B} proposed neural networks as \emph{ansatzs} for
the time evolution of the chemistry and trained them using the ODEs in the loss
function. Along similar lines, \cite{branca2024emulating} used Neural Operators to emulate the 
chemical network \citep{DBLP:journals/natmi/LuJPZK21}.

Despite the amount of work done in the field, we think our architectural decisions, training strategy, 
and size of the chemical network 
can be of impact in the community. Moreover, we have trained this emulator to explore the impact of the
sulphur elemental abundance on the physics and chemistry of the
interstellar medium. In this paper, we focus on the gas ionization degree, an essential parameter for the dynamics of interstellar clouds.

\section{Chemical model}
We performed the abundance calculations using Nautilus 1.1 \citep{Ruaud2016}, a 
three-phase model, in which gas, grain surface and grain mantle phases, and their interactions, 
are considered. It solves the kinetic equations for the gas-phase and the solid species at the 
surface of interstellar dust grains.
Species from the gas-phase can physisorb at the surface of dust grains with an energy that depends 
on the binding energy of surface species \citep{Wakelam2017}. They can be released to the gas phase 
by thermal desorption \citep{Hasegawa1992},  grain heating induced by cosmic-rays 
(following \citealp{HasegawaHerbst1993}, impact of UV photons (photodesorption, \citealp{Ruaud2016}), 
exothermicity of surface reactions (chemical desorption, \citealp{Minissale2016}), and sputtering by 
cosmic-rays \citep[following][]{Wakelam2021}. Detailed descriptions of the non-thermal desorption 
processes included in Nautilus can be found in \citet{Wakelam2021}. The ice surface phase is composed 
of the first few monolayers of species on top of the grains (four in our case), while the rest of the 
molecules below these surface layers represents the bulk of the ice. The refractory part of the grains 
(below the bulk) is chemically inactive. The species on the surface can desorb into the gas phase, 
contrary to the species in the bulk. Only sputtering by cosmic-rays can directly desorb species from 
the bulk ice. The version used in this paper allows to calculate the dust temperature using the 
expression of \citet{Hocuk2017} and the cosmic ray ionization rate per H$_2$ molecule ($\zeta$) as 
a function of A$_V$ following the expresion of \citet{NeufeldWolfire2017} as described in \citet{Clement2023}.
However, we have disabled these options to create our database since we were interested in determining the 
variation of the chemical abundances with these parameters.

\section{Neural emulator}
Learning emulators for computationally heavy numerical procedures 
via neural networks is a well-established technique in the field of machine learning.
The idea exploits the well-known fact that neural networks are universal function approximators \citep{cybenko88}
to train them to learn the mapping between the input 
(physical parameters) and the output (the result of the numerical procedure).
In theory, they can learn the mapping between input and output spaces with arbitrary precision
provided that the network is large enough and the training set is sufficiently rich. In practice,
this is often limited and only a limited precision is reachable.

In our model, the solution to the kinetic equations give the abundance of a set of 192 species
as a function of time $t$. The abundance is also dependent on the external physical
conditions, that are characterized with the following parameters: the temperature ($T$), 
the initial gas density ($n_H$), the cosmic ray ionization rate ($\zeta_{\mathrm{H}_2}$), the sulfur abundance ($\left[S\right]$)
and the scaling factor for the surrounding UV flux ($\chi$). As a consequence, our aim
is to end up with a neural network $A(t|\mathrm{M},\vect \theta)$
that approximates the abundance for molecule M at time $t$, given the physical conditions
summarized in the vector $\vect \theta=(T,n_\mathrm{H},\zeta_{\mathrm{H}_2},\left[S\right],\chi)$.

\subsection{Architecture}
A crucial ingredient to end up with a succesful emulator is to find a suitable architecture.
We propose to emulate the interstellar chemistry using a neural field (NF), also
known as implicit neural representations (INR) or coordinate-based representations 
(CBR) \citep[see][]{mildenhall2020nerf,2022arXiv221114879B,2022AGUFMSH45D2368J,2023SoPh..298..135A}. 
These are fully connected (FC) neural networks that are used to map  
coordinates on the space (or space-time) to coordinate-dependent field quantities. NFs have
desirable properties for serving as emulators over other parametrizations. They 
are very efficient in terms of the number of free parameters. Most important, they 
produce continuous (and differentiable\footnote{Or, more precisely, sub-differentiable in case of using
activation functions that are not differentiable everywhere.}) fields, which can
then be evaluated at arbitrary inputs. Finally, they have a strong 
implicit bias, favoring specific signals.

In our case, we use a NF in which the input is the time at which we want
to predict the abundance of a specific species and the output is the abundance
of that species. Obviously, the NF needs to be conditioned on the physical parameters of the cloud
and on the specific species. We show later how we do this conditioning technically. 
A typical NF gives the abundance at time $t$ of a certain species M and for a set of
physical conditions $\vect \theta$ by the following function composition:
\begin{equation}
    A(t|M,\vect \theta) = \phi_{N}^{\mathrm{M},\vect\theta} \circ \cdots \circ \phi_1^{\mathrm{M},\vect\theta} \circ \phi_0^{\mathrm{M},\vect\theta}(t),
    \label{eq:nf1}
\end{equation}
where we explicitly state that each layer of the NF is conditioned on the
vector $\vect\theta$, that contains information about the physical conditions and the species, 
and the specific species to be predicted, M.
Note that $(f \circ g)(\mathbf{x}) = f(g(\mathbf{x}))$ refers to the function composition operation.
Typically, each layer of the NF is given by:
\begin{equation}
   \phi_i^{\mathrm{M},\vect\theta}(\mathbf{x}) = \sigma(\mathbf{W}_i \mathbf{x} + \mathbf{b}_i),
\end{equation}
where $\mathbf{W}_i$ and $\mathbf{b}_i$ are weight matrices and bias terms, respectively, while
$\sigma$ is a non-linear activation function. In our case, we choose the rectified linear 
unit \citep[ReLU;][]{relu10} as the activation function.

\begin{figure}
   \centering
   \includegraphics[width=\columnwidth]{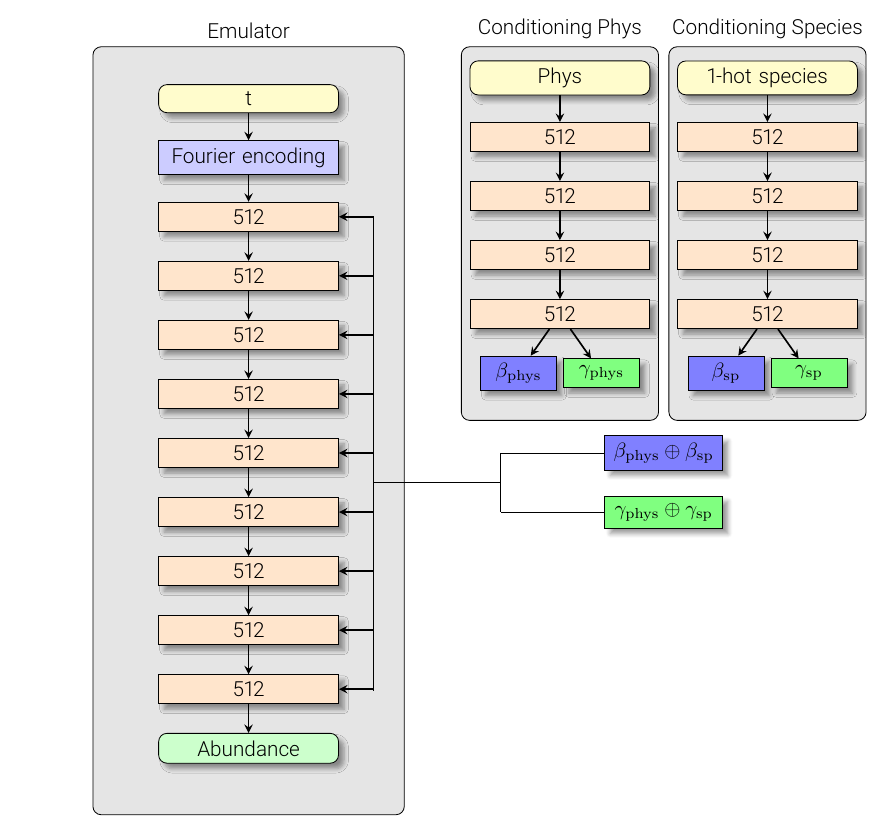}
   \caption{Architecture used in the emulator. The ``Emulator'' block encapsulates the NF that predicts
   the abundance of a certain species at a given time. The two conditioning blocks produce
   vectors that are concatenated together and used in the neural field. Yellow blocks are input
   layers. Orange blocks are fully connected layers, with the number of neurons indicated for
   clarity. Light purple color refer to the Fourier encoding layer explained in Eq. (\ref{eq:fourier_encoding}). The conditioning vectors enter into the fully connected layers following the FiLM paradigm.\label{fig:architecture}}
   \end{figure}

The conditioning is done via two FC neural networks, one that deals with the physical parameters
and the other one with the species. The input of the first neural network is the vector
$\vect \theta$, all of them
properly normalized as described in Section \ref{sec:dataset}. The input 
of the second neural network is a representation of the molecule in a 1-hot
encoding\footnote{Another option is to directly learn vector embeddings for all species, instead
of them being computed from 1-hot vectors with the aid of neural network.} (i.e., a vector of dimension equal to the number of species, with a 1 in the
component corresponding to the species and 0 in the rest). 
The outputs of each one of these neural networks are two scaling vectors, 
$\vect{\gamma}^{\vect \theta}_\mathrm{phys}$ and $\vect{\gamma}^\mathrm{M}_\mathrm{sp}$, and two
bias vectors, $\vect{\beta}^{\vect \theta}_\mathrm{phys}$ and $\vect{\beta}^\mathrm{M}_\mathrm{sp}$. 
We follow the ideas behind feature-wise linear modulation \citep[FiLM;][]{film17}, so that 
each layer of the NF is modified using the previous vectors according to:
\begin{equation}
   \phi_i^{\mathrm{M},\vect\theta}(\mathbf{x}) = \sigma\left[ 
      \vect{\gamma}^{\mathrm{M},\vect\theta} (\mathbf{W}_i \mathbf{x} + \mathbf{b}_i) + \vect{\beta}^{\mathrm{M},\vect\theta} \right],
\end{equation}
where $\vect{\gamma}^{\mathrm{M},\vect\theta} = \vect{\gamma}^\theta_\mathrm{phys} \oplus \vect{\gamma}^\mathrm{M}_\mathrm{sp}$
and $\vect{\beta}^{\mathrm{M},\vect\theta} = \vect{\beta}^\theta_\mathrm{phys} \oplus \vect{\beta}^\mathrm{M}_\mathrm{sp}$
are obtained via concatenation ($\oplus$ is the vector concatenation operator).
One of the advantages of the FiLM approach to conditioning is that it is very
easy to implement and it does not require a significant modification of the NF architecture.
Additionally, all layers of the NF receive conditioning information, which 
facilitates the adaptation of the NF to the complex dependence of the output
on the physical conditions.

\begin{figure*}
    \centering
    \includegraphics[width=\textwidth]{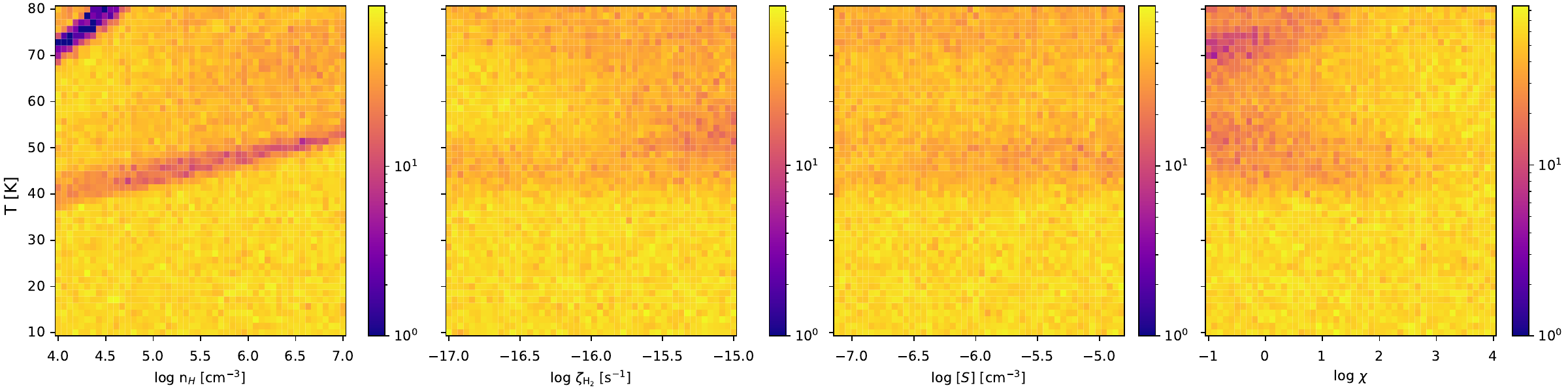}
    \caption{Two-dimensional histogram showing the covering of all 126k models of the training set
    shown as combinations of temperatures and the rest of parameters.\label{fig:training_set}}
    \end{figure*}
    
NFs, with the previous definitions, are known to suffer from the so-called spectral bias 
\citep{pmlr-v97-rahaman19a,WANG2021113938}, which prevents them from learning high-frequency functions.
We follow \cite{tancik2020fourier} and solve this problem by first passing the time
coordinate through a Fourier feature mapping that increases the dimensionality from 
1 to $2m$ (with $m$ the number of Fourier features):
\begin{equation}
   A(t|M,\vect \theta) = \phi_{N}^{\mathrm{M},\vect\theta} \circ \cdots \circ \phi_1^{\mathrm{M},\vect\theta} \circ \phi_0^{\mathrm{M},\vect\theta}(\gamma(t)),
   \label{eq:nf2}
\end{equation}
where
\begin{equation}
   \gamma(t) = \left[t,\cos(2 \pi \mathbf{B} t), \sin(2 \pi \mathbf{B} t) \right]^\dag.
   \label{eq:fourier_encoding}
\end{equation}
We also include the time coordinate itself, to allow for outputs with 
very low-frequency time dependencies.
The matrix elements of the matrix $\mathbf{B} \in \mathbb{R}^{m \times 1}$ are randomly sampled from a normal distribution
with zero mean and standard deviation $\sigma_F$. After a non-exhaustive parameter
search, we choose $\sigma_F=2$ and $m=512$, which produces a Fourier feature mapping
of dimensionality 1025. 

Although we have found success with adding Fourier features and using ReLU activation
functions, other solutions like the \texttt{SIREN}s architecture 
\citep[Sinusoidal Representation Networks;][]{sitzmann2020implicit}, which uses 
periodic activation functions, can also be found in the literature and could be
explored for this problem in the future.

\begin{figure}
\centering
\includegraphics[width=\columnwidth]{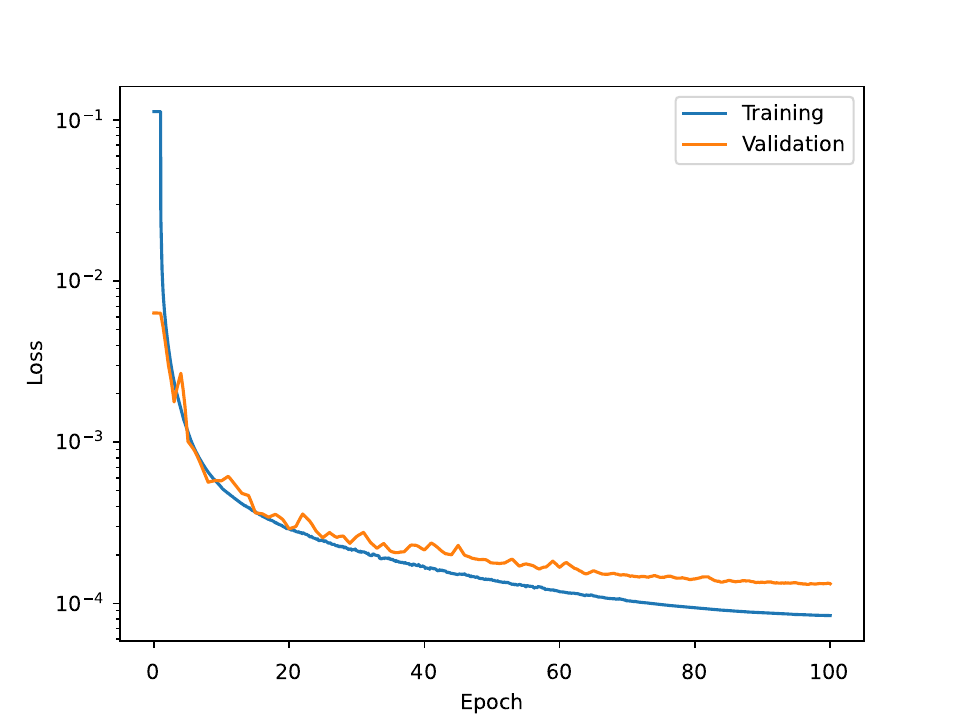}
\caption{Evolution of the training and validation losses during training. An exponential
average with a window size equal to the number of batches in an epoch has been applied.\label{fig:loss}}
\end{figure}

The graphical representation of the final model is depicted in 
Fig. \ref{fig:architecture}. The two conditioning
neural networks contain three hidden layers with 512 neurons. They produce the
two conditioning vectors $\vect{\gamma}_i$ and $\vect{\beta}_i$ with 256 elements each.
After concatenating them, they are used in the NF emulator to condition every 
hidden layer of the NF with 512 neurons. The final output is a single number, the abundance
of the molecule for the specified physical conditions. The total
number of trainable parameter is 4.83M, with 2.62M associated with the NF, and 1.05M and 1.15M
associated with the conditioning networks for the physical conditions and the species, respectively.

Thanks to the interpolation capabilities of the NF, we can predict the abundance of any molecule
at any arbitrary time between 1 and $10^7$ yr. Additionally, given that the NF is sub-differentiable,
one can seamlessly compute derivatives of the abundance of any arbitrary species with respect to both the physical 
conditions and time. This can be of great potential interest for the analysis of the chemistry and
also for using the chemistry emulator in other downstream tasks.

\begin{table}
   \centering
   \caption{Physical conditions considered in the emulator, together with the range
   of values and the transformation used for sampling them.}
   \label{tab:phys}
   \begin{tabular}{l|ccc}
   \hline
   \hline
   Parameter & Minimum & Maximum & Sampling \\
   \hline
   $t$ [yr] & 1 & 10$^7$ & log \\
   $T$ [K] & 10 & 80  & linear \\
   $n_H$ [cm$^{-3}$] & 10$^4$ & 10$^7$ & log \\
   $\zeta_{\mathrm{H}_2}$ [s$^{-1}$] & 10$^{-17}$ & 10$^{-15}$ & log \\
   $\left[S\right]$ [cm$^{-3}$] & 7.5$\times 10^{-8}$ & 1.5$\times 10^{-5}$ & log \\
   $\chi$ & 0.1 & 10$^4$ & log \\
   \end{tabular}
   \end{table}

\subsection{Training set}
\label{sec:dataset}
The training set is built by solving the interstellar chemistry using Nautilus for a set 
of physical conditions and a total of 192 species. We add atomic and molecular
species, as well as ions and ices. The chemistry network is solved for
64 times equispaced in log space between 1 and 10$^7$ yr. A total of 150k models are
run for building the training set. Roughly 16\% of these models do not pass
a hard timeout threshold of 5 min and are then discarded\footnote{This hard threshold is applied
to allow computing the database in a reasonable time. Even with this threshold, the calculation
requires $\sim4$ hours in 512 cores.}. This leaves us with 126k models, which is still
a sufficiently large amount of models for training. The total number of training examples
is, therefore $\sim$126k$\times$192, which amounts to 24.2M. The space covering of the 
training models is depicted in Fig. \ref{fig:training_set}, where we witness a region of
a reduced number of models for low densities and high temperatures, where Nautilus has
problems converging the kinetic equations.

\begin{figure*}
    \centering
    \includegraphics[width=\textwidth]{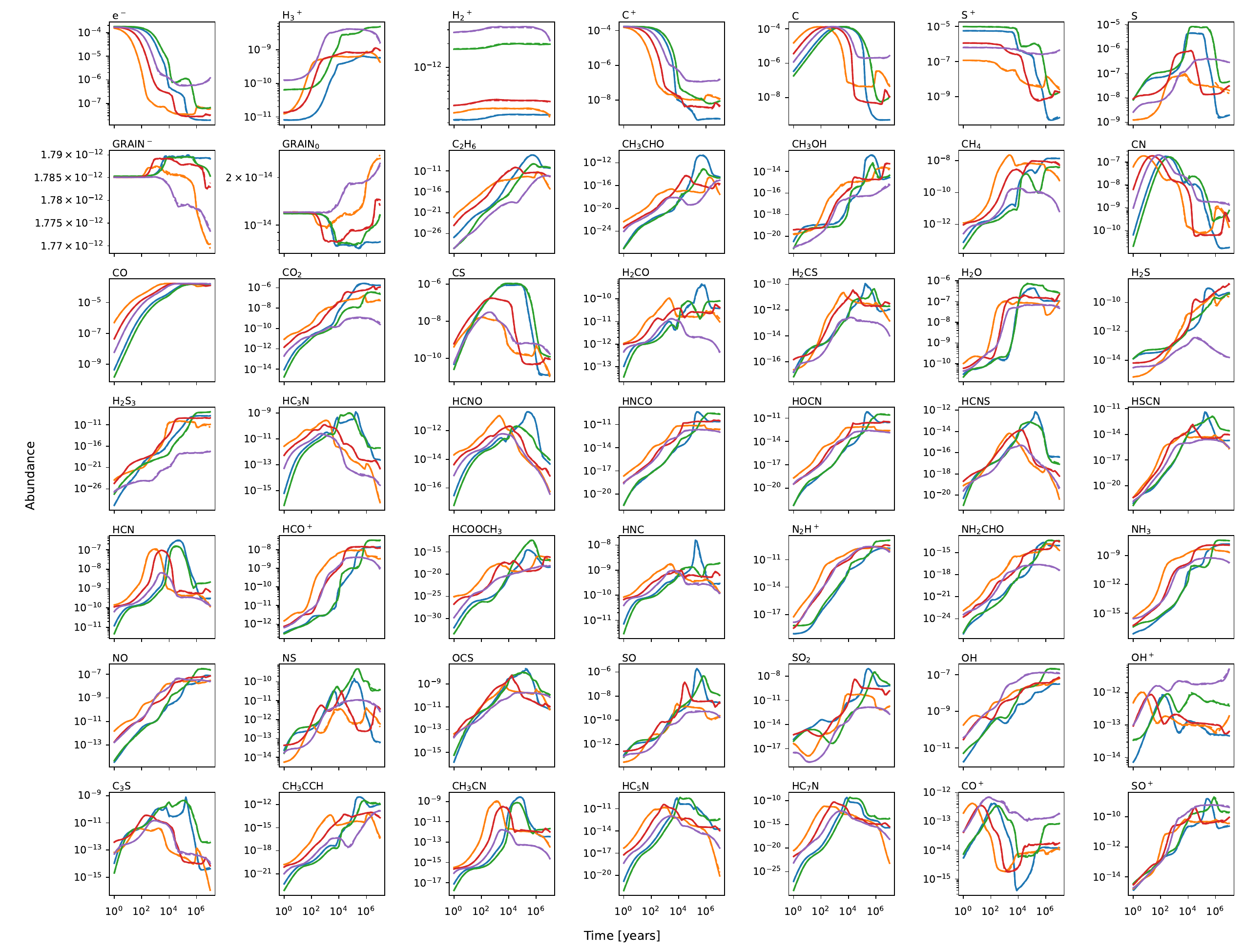}
    \caption{Five samples per species of those considered of more interest. We
    show the time evolution of the abundance computed with the neural emulator
    in solid line, while the original time evolution of the test set is 
    shown in dashed lines. These have been selected randomly with no cherry-picking.
     \label{fig:samples}}
    \end{figure*}
    
The physical conditions are selected randomly
in the intervals defined in Table \ref{tab:phys}. The temperature is sampled uniformly
without transformation, while the rest of the parameters are sampled uniformly in log space
given that they are positive quantities and span several orders of magnitude. A subset of 10\% of the models is
left out of the training set and used as a validation set. A final set of 5k models
is obtained for testing purposes and it is used to compute the final performance of the
emulator. We point out that other methods to sample the physical conditions, such as Latin 
hypercube sampling, could have been used, but we have found that random sampling is 
sufficient for this problem. Additionally, it allows us to seamlessly extend
the training set outside the currently considered regions.

The input physical conditions (in linear or log scales) are normalized to the interval $[-1,1]$
when used as input to the conditioning neural networks. Likewise, $t$ is also normalized to the
same interval after being transformed to log scale. The output abundances are similarly treated in log scale.
However, since a large variability is found in the abundances, we normalize each time bin and
molecule separately. This normalization is stored and applied in reverse when the model
is used in production.


\subsection{Training}
The training proceeds by minimizing the mean squared error between the output of the
NF and the abundance for all considered  molecules. 
The initial learning rate is set to 3$\times 10^{-4}$ and it is slowly reduced
to 3$\times 10^{-5}$ using a cosine annealing schedule \citep{loshchilov2017sgdr}. The optimization is
carried out using the Adam optimizer \citep{Kingma2014} with a batch size of 1024 for 100 epochs.
We utilize an NVIDIA GPU 4090 with 24 GB of memory. Given that the training takes around 24 hours,
we did not do an exhaustive hyperparameter search. The evolution of the training and 
validation losses during training are displayed in Fig. \ref{fig:loss}, showing no
evidence of overtraining. We found the saturation of the validation loss but with a tiny
difference with respect to the training loss. Although after epoch $\sim 40$ the
decrease in the validation loss is relatively slow, we preferred to continue training for 100 epochs.
The reason is that the diagnostics shown in the following sections carried out during intermediate
epochs showed that the training was still improving.

\begin{figure*}
   \centering
   \includegraphics[width=\textwidth]{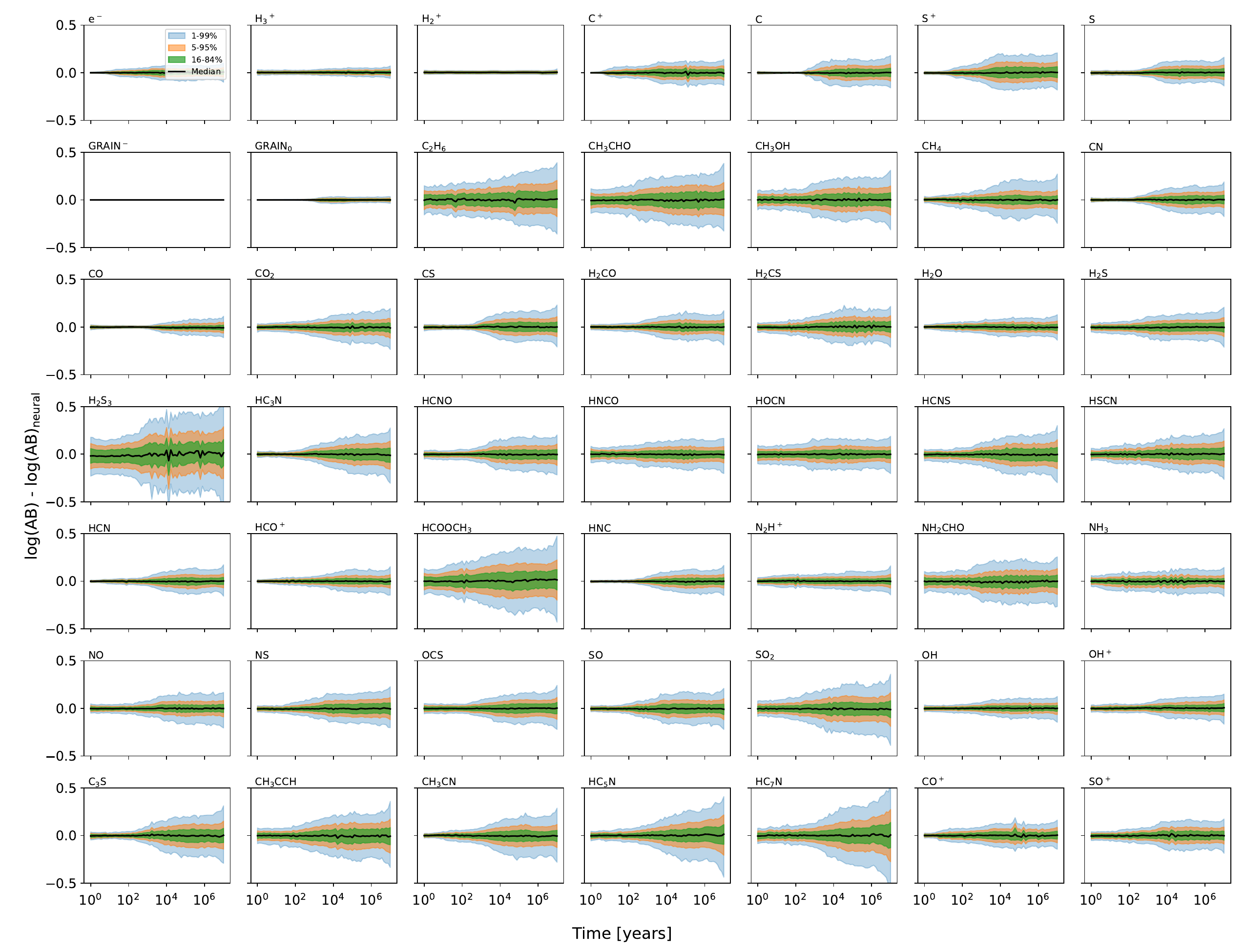}
   \caption{Difference in logarithm between the emulator and the original run
   for selected species. The solid line shows the median difference, while 
   the shaded area shows the quantiles 1-99 in blue (approximately $3\sigma$), 
   5-95 (approximately $2\sigma$) in orange and 16-84 (approximately $1\sigma$)
   in green.
    \label{fig:stats}}
   \end{figure*}

In order to adapt the complexity of the NF to the specific problem, we opt for starting
the optimization taking only into account the low frequency components in the Fourier feature 
mapping, while gradually turning on higher frequency components during training. To this end, we extend the approach used by 
\cite{lin2021barf} as follows. We reorder the rows of the matrix $\mathbf{B}$ in terms of
their $\ell_2$ norm, with low frequencies appearing first. We then weight the Fourier 
features as follows:
\begin{equation}
   \gamma(t) = \left[t,\vect w(\alpha) \cos(2 \pi \mathbf{B} t), \vect w(\alpha) \sin(2 \pi \mathbf{B} t) \right]^\dag,
\end{equation}
where the $k$-th element ($k=0,1,\ldots,m-1$) of the weighting function is given by
\begin{equation}
   w_k(\alpha) = 
   \begin{cases}
      0 &\quad\text{if } \alpha < k\\
      \frac{1 - \cos((\alpha - k) \pi)}{2} &\quad\text{if } 0 \le \alpha - k < 1 \\
      1 &\quad\text{if } \alpha - k \ge 1,
    \end{cases}
\end{equation}
 and $\alpha \in [0, m]$ is slowly increased during training from 0 to $m$.
We choose to linearly increase $\alpha$ from 0 to $m$ during the first 25000 batches, and set $\alpha=m$ (no
weighting) for the rest of the training. This procedure allows the NF to start learning low frequencies
first and then gradually incorporate high frequencies. We have found that this procedure gives 
more consistent and faster results during training than simply starting with all 
frequencies active in the Fourier encoding from the beginning.

\section{Validation}
Once the emulator is trained, we evaluate the performance using the 5k Nautilus models
computed specifically for this purpose. Figure \ref{fig:samples} shows the time
evolution of five randomly selected models. We only display a subset of species of 
certain relevance, although the output is obtained for all 192 species. The time 
evolution of the abundance computed with the neural emulator is displayed
with solid lines, while the time evolution computed with Nautilus is shown 
with dashed lines. A simple visual comparison demonstrates that the emulator is able to
correctly reproduce the time evolution of these species.

The neural emulator provides a significant gain in computing time. A typical Nautilus model
requires of the order of 120-240 s to produce the output for all considered molecules at 64 timesteps equispaced in
logarithm between 1 and 10$^7$ yr. This number fluctuates depending on the physical
conditions, which change the stiffness of the kinetic equations and impacts the
total computing time. Our emulator is able to compute 4.8 models/s in
an Intel Xeon CPU E5-2660@2.60 GHz, a rather standard desktop CPU from 2017. We clarify
that a ``model'' evaluation for the emulator requires using the emulator for all 192 species at 64 timesteps.
This number goes up
to 87 models/s when using a GPU if models are computed one by one. This does not saturate the
computing power of the GPU, which are specially tailored to compute many models in parallel. This 
saturation happens, in our specific conditions, when $\sim$32 models are evaluated simultaneously. At this 
rate, we are able to compute 146 models/s. Therefore, our neural emulator is between $10^4$ and
$3\times 10^4$ times faster than Nautilus. Additionally, the computing time is not sensitive to the specific
physical conditions, carrying out exactly the same number of operations for the whole
parametric volume defined in Tab. \ref{tab:phys}.


An arguably better representation of the performance of the emulator is given in Fig. 
\ref{fig:stats}, together with Figs. \ref{fig:stats1}, \ref{fig:stats2} and \ref{fig:stats3}
in Appendix A. We display the distribution of differences in logarithm between the abundance
predicted by the emulator and the original one computed with Nautilus. For simplicity, we 
characterize the distributions showing the percentiles. The black solid line shows the median difference, which is
interestingly close to zero for all species. This means that deviations above
and beyond are equally likely. This is confirmed with the shaded areas, which display
the range between percentiles 1 and 99 in blue, 5 and 95 in orange, and 16 and 84
in green. The figure shows that the
difference lies below 0.2 dex in the majority of species of relevance for all
times with 98\% probability. Only for a few species (like H$_2$S$_3$, HC$_7$N, \ldots)
this difference goes up to 0.4 dex for the latest phases of the chemical evolution.
Note that we also show species in the ices (characterized with the letter
J at the beginning when the species is in the surface, and K for those in the ice bulk).


The difference is consistently very small for the initial 
phases of the evolution for all species. The reason is that the initial abundance for all species
is fixed for all models so it is easy to be learned by the emulator. The specific final abundance
depends on the time evolution of the chemistry, which turns out to be more
difficult to learn. The abundance at very large times converges
to the equilibrium value, which should only depend on the physical conditions.
Despite this fact, we find that the uncertainty in the emulator is often larger
in the final stages of the evolution, although still consistently below 0.4 dex. This is because the equilibrium value
is often reached at very large times so the uncertainty in the emulator
is larger. 


\begin{figure*}
        \centering
        \includegraphics[width=\textwidth]{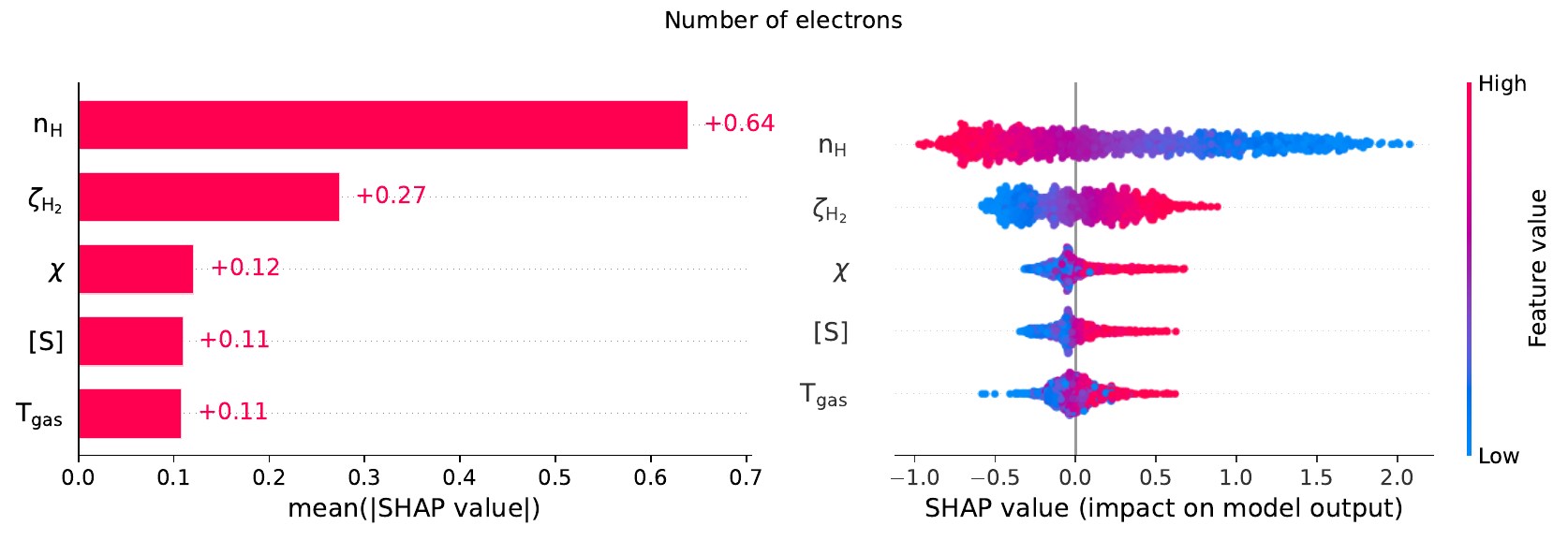}
        \caption{Plots summarizing the SHAP values and feature importance for the 2048 predictions of the number of electrons. \emph{Left:} feature importance as defined in Eq. \ref{eq:importance} for each feature of the emulator. \emph{Right:} a so-called beeswarm plot where dots show the SHAP values of each prediction for all features, colored depending on their feature value.}
        \label{fig:rankBeeSwarm}
\end{figure*}

\section{Importance of the ionization degree and initial sulphur abundance}
Neural networks are often described as black boxes in the sense that, even 
though they can approximate the results well, the structure of the neural network 
itself generally does not provide any insight into the model being approximated. It is a reality
(although the machine learning community is actively working towards solving this issue) that it is hard
to understand the learned relationship between the input parameters and the outcome of the model. This information 
is sometimes of great interest in astrochemistry, as questions like what physical conditions favor or hinder 
the production of a given chemical species are frequently faced. 

A potential avenue to understand the relationship between inputs and outputs in astrochemical models was 
addressed in \citet{Heyl2023} using Shapley values \citep{Shapley1953}. 
The Shapley value is an approach from game theory that provides a way to fairly distribute the outcome 
of a cooperative game among the players. In the context of astrochemical emulation, the chemical emulator is 
the cooperative game, the outcome (or prediction) is the abundance of a given chemical species, and the players, 
commonly referred as features, are the input parameters of the emulator. Shapley values therefore allow us to quantify 
the importance of the features for a given prediction. In the following, we give a brief description of 
the method applied to the chemical emulator. For a full description of the Shapley value approach, we refer 
to \citet{Shapley1953}. 

Let us define $N=\{T,n_\mathrm{H},\zeta_{\mathrm{H}_2},\left[S\right],\chi\}$ as the full set of input parameters of the
neural emulator and $\vect \theta^{j}$ as the vector of input parameters of training example $j$.
Under the Shapley value framework \citep{Lundberg2017,Heyl2023}, the explanation of the prediction $f(\vect \theta^{j})$ is
assumed to be given by the following linear model:
\begin{equation}\label{eq:additiveExplanation}
    f(\vect \theta^{j}) \sim \langle f \rangle + \sum_{i=1}^{|N|}\phi^j_{i},
\end{equation}
where we made the notational simplification $f(\vect \theta^{j})=A(t|\mathrm{M},\vect \theta^{j})$ once
$t$ and M are fixed.
Put in words, the simplified explanation model is built by adding $\langle f \rangle$, the average value of the prediction, together
with the Shapley values $\phi^j_{i}$ associated with all the $i$-th input parameter at the $j$-th training example.
This linear model allows one to extract the importance of each input parameter for a given prediction by
simply comparing the Shapley values.

Shapley values can be obtained from the predictions of our model. Let
$S \subseteq N$ be a subset of $N$ (e.g., $S=\{T,n_\mathrm{H}\}$). The vector $\vect \theta^{j}_{S}$ is obtained by keeping the elements
from the subset $S$ as they are and setting the rest of parameters to a random value extracted from 
the data set. With these definitions, the Shapley value 
$\phi^j_{i}$ associated with the input parameter $i$ and training example $j$ is the weighted average of the marginal 
contribution of the feature $i$ over all possible coalitions of features excluding the feature $i$:
\begin{equation}
    \phi^j_{i} = \sum_{S\subseteq N\setminus\{i\}}\frac{|S|!\ (|N|-|S|-1)!}{|N|!}
    \left[f\left(\vect \theta^{j}_{S\cup\{i\}}\right)-f\left(\vect \theta^{j}_{S}\right)\right].
\end{equation}
As indicated, the summation is carried out for all subsets that do not include the 
input parameter of interest. Note that $|N|$ and $|S|$ are the number of elements in sets $N$ and $S$, respectively.

Once the Shapley values are computed, the global importance of feature $i$ in the prediction, denoted as $I_{i}$, 
is given by the mean absolute value of the Shapley values for that feature over the entire dataset:
\begin{equation}\label{eq:importance}
    I_{i} = \frac{1}{d} \sum_{j=1}^d |\phi^j_{i}|,
\end{equation}
where $d$ is the number of examples in the dataset.
Obviosuly, the relative importance of feature $i$ can be obtained as:
\begin{equation}\label{eq:relImportance}
\bar{I}_{i} = \frac{I_{i}}{\sum_{j} I_{j}}.
\end{equation}

%

\begin{figure*}
    \centering
    \begin{subfigure}[b]{0.246\textwidth}
        \centering
        \includegraphics[width=\textwidth]{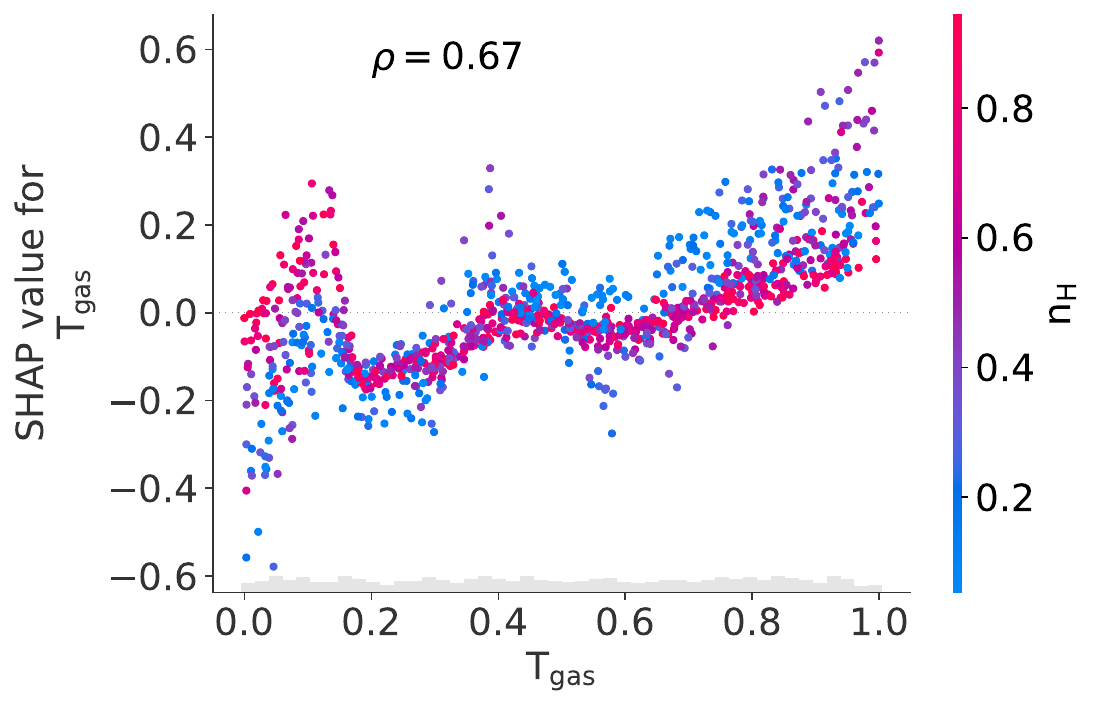}
    \end{subfigure}
    \begin{subfigure}[b]{0.246\textwidth}
        \centering
        \includegraphics[width=\textwidth]{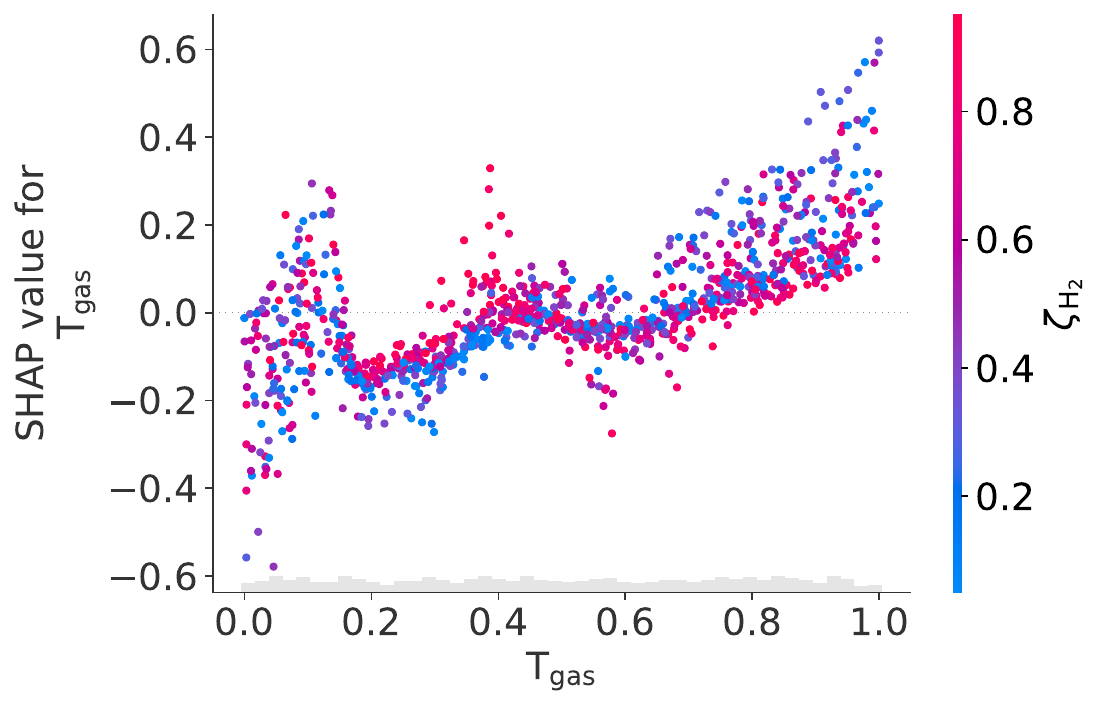}
    \end{subfigure}
    \begin{subfigure}[b]{0.246\textwidth}
        \centering
        \includegraphics[width=\textwidth]{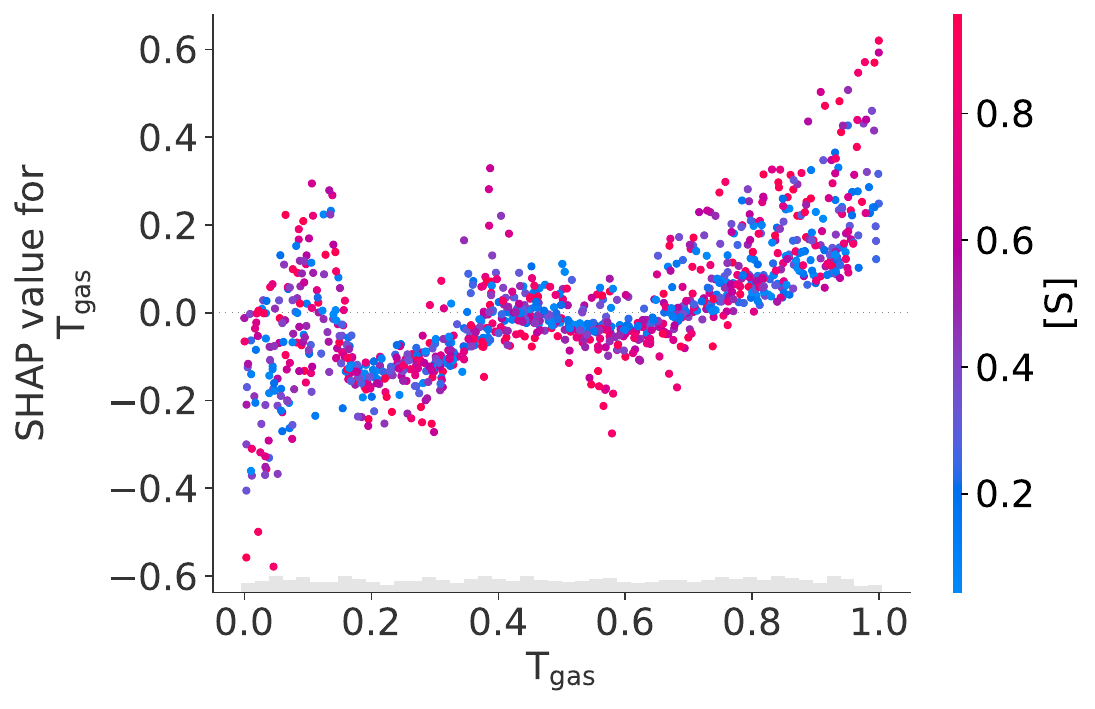}
    \end{subfigure}
    \begin{subfigure}[b]{0.246\textwidth}
        \centering
        \includegraphics[width=\textwidth]{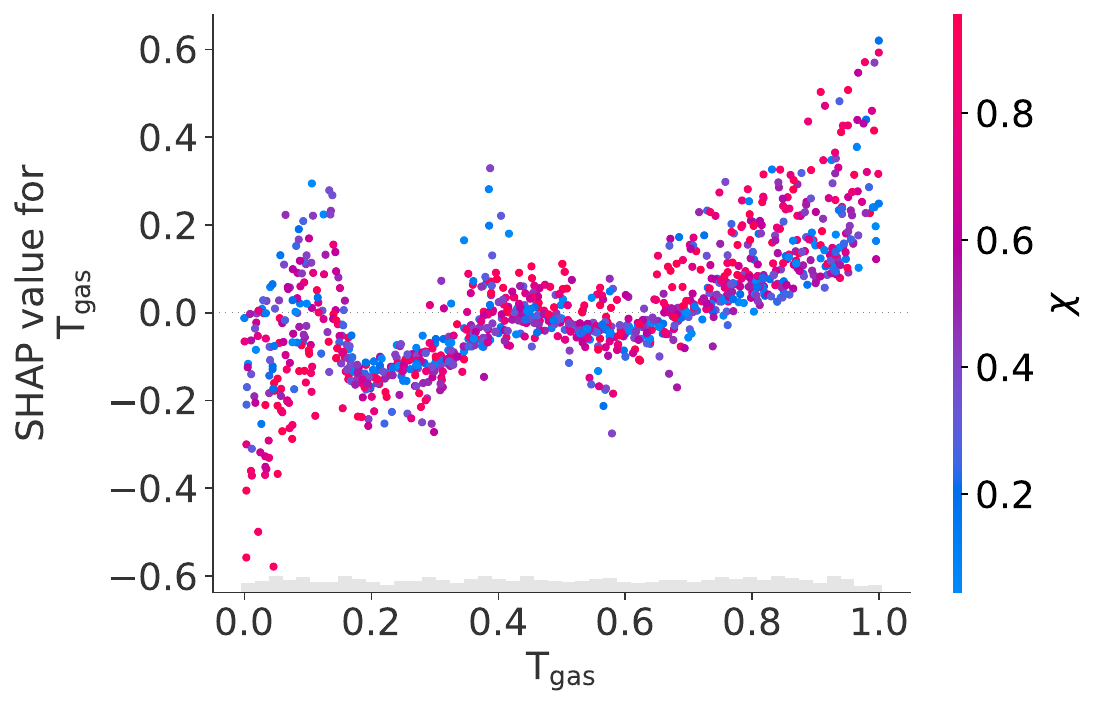}
    \end{subfigure}
    ~
    \begin{subfigure}[b]{0.246\textwidth}
        \centering
        \includegraphics[width=\textwidth]{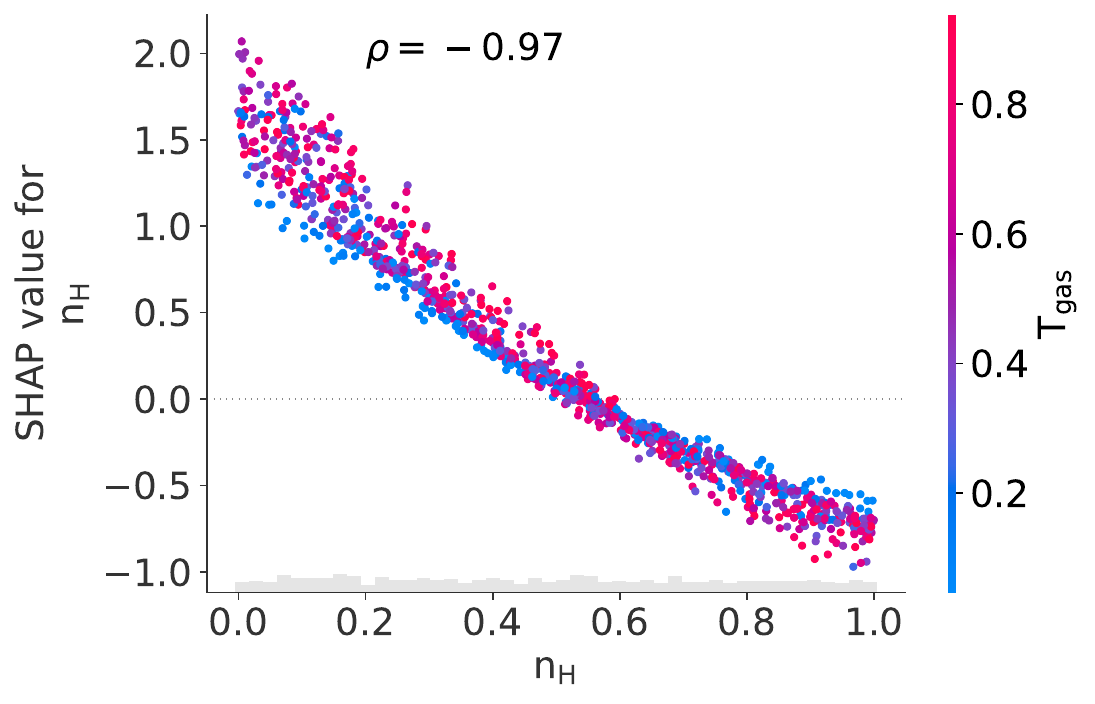}
    \end{subfigure}
    \begin{subfigure}[b]{0.246\textwidth}
        \centering
        \includegraphics[width=\textwidth]{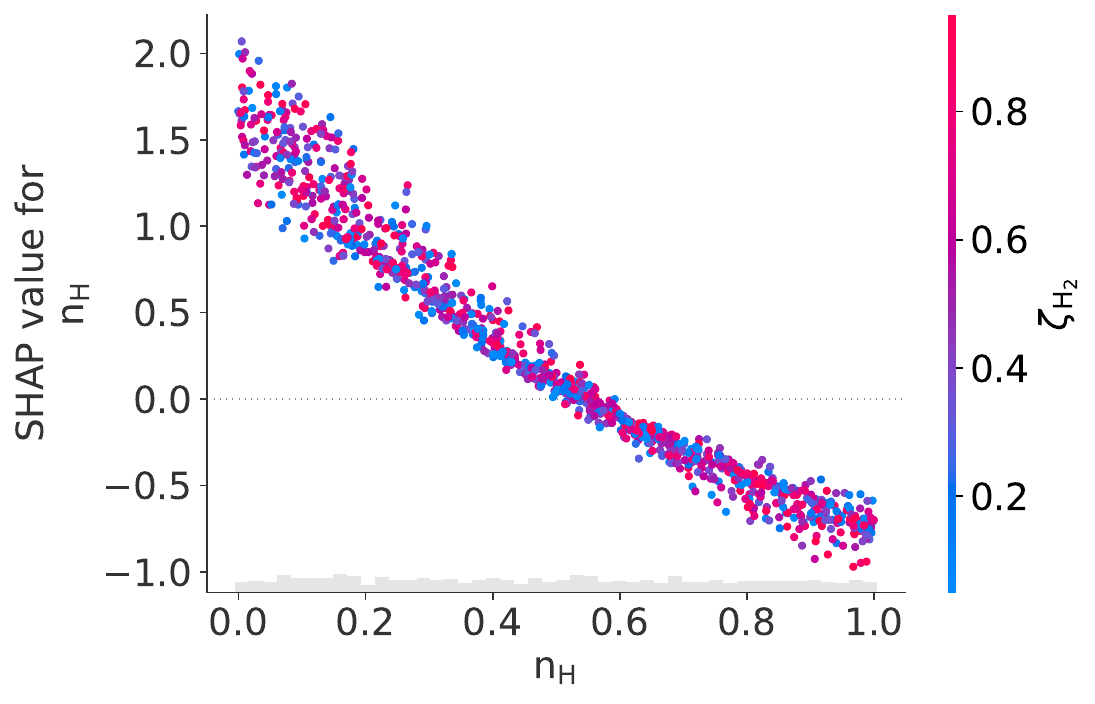}
    \end{subfigure}
    \begin{subfigure}[b]{0.246\textwidth}
        \centering
        \includegraphics[width=\textwidth]{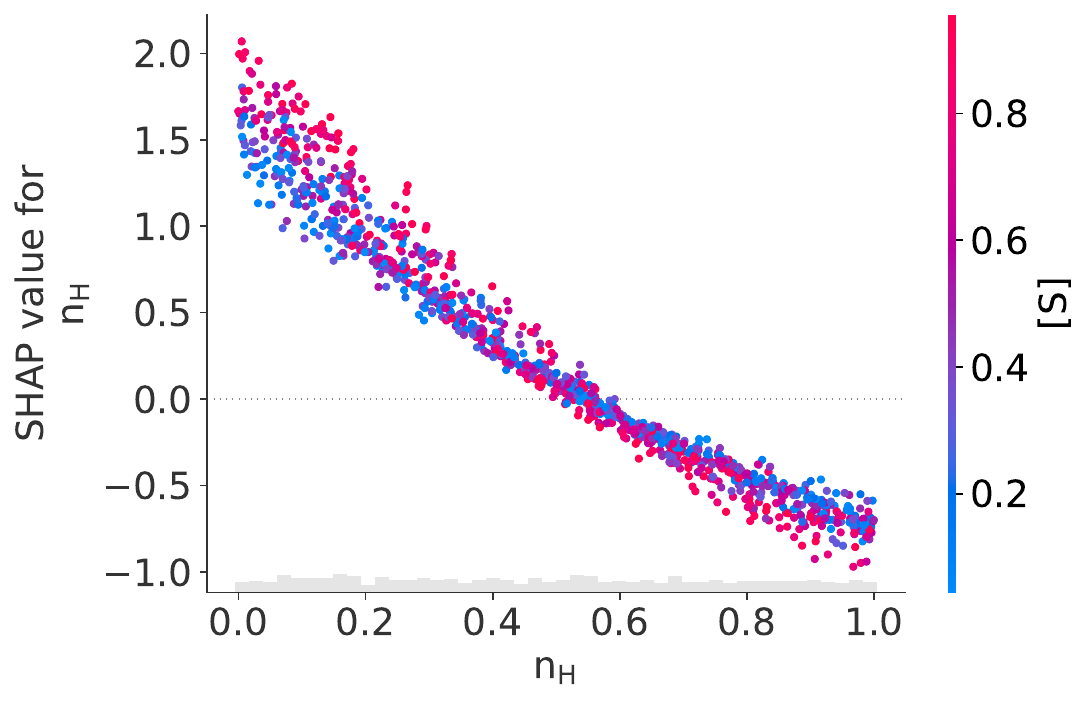}
    \end{subfigure}
    \begin{subfigure}[b]{0.246\textwidth}
        \centering
        \includegraphics[width=\textwidth]{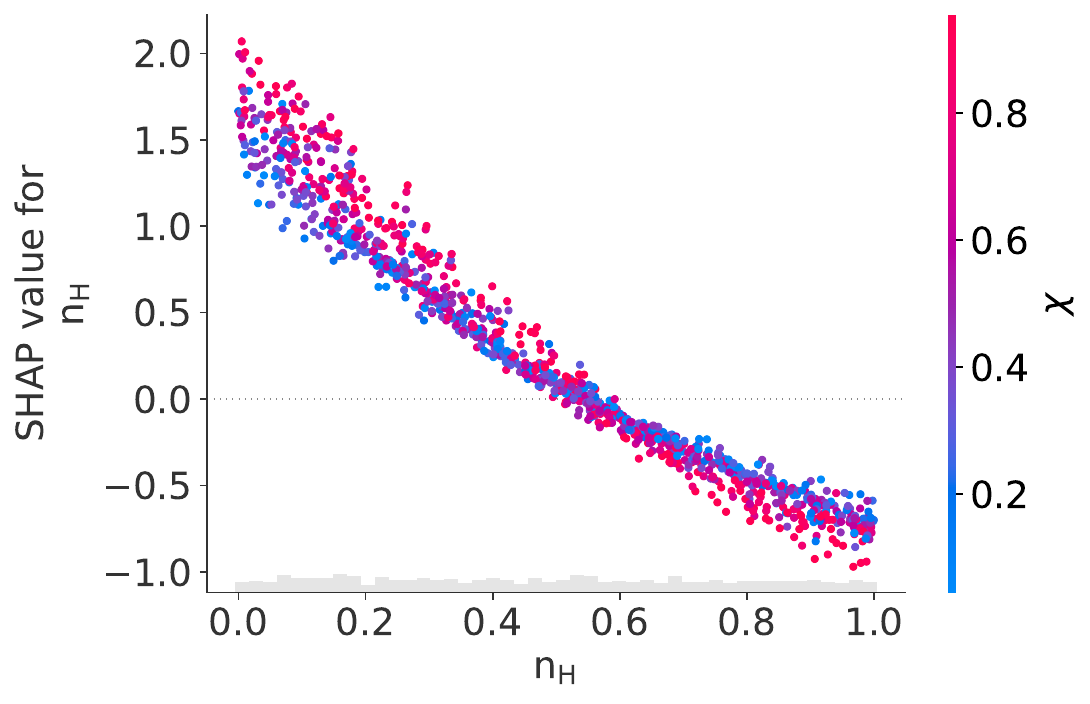}
    \end{subfigure}
    ~
    \begin{subfigure}[b]{0.246\textwidth}
        \centering
        \includegraphics[width=\textwidth]{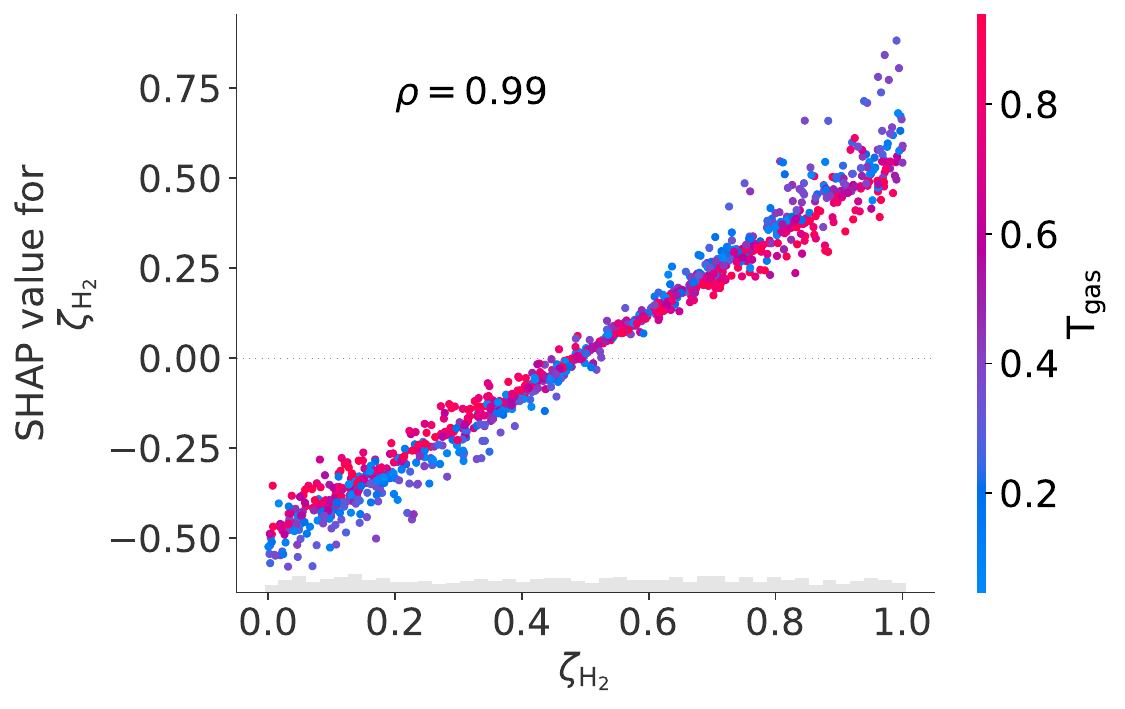}
    \end{subfigure}
    \begin{subfigure}[b]{0.246\textwidth}
        \centering
        \includegraphics[width=\textwidth]{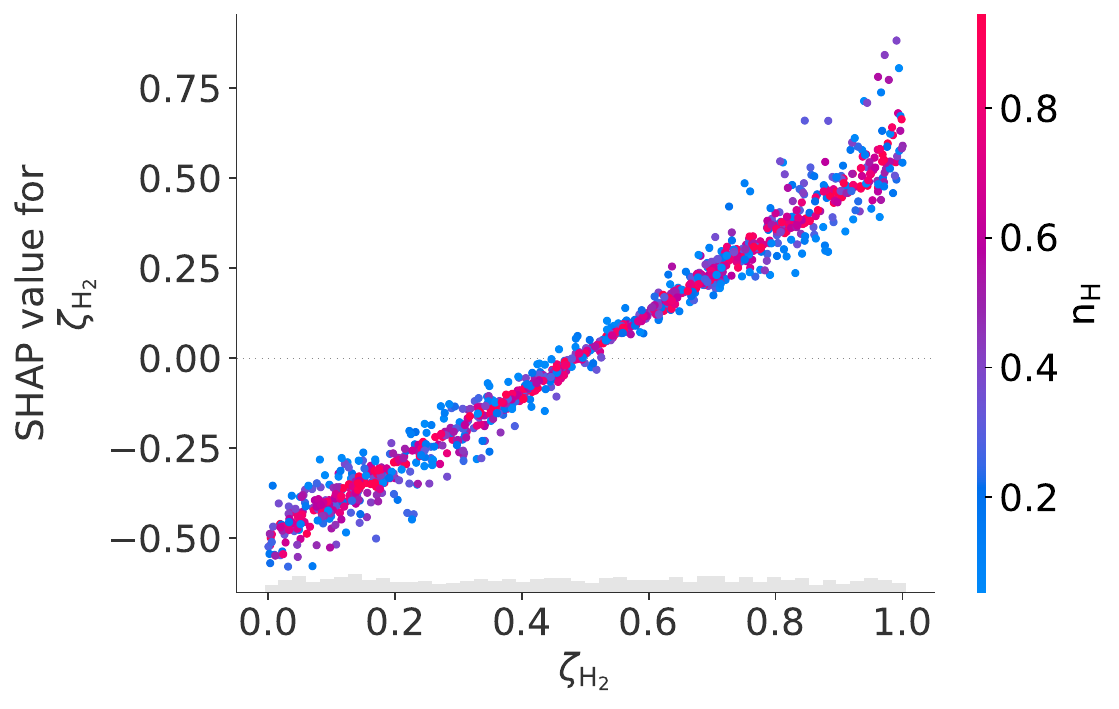}
    \end{subfigure}
    \begin{subfigure}[b]{0.246\textwidth}
        \centering
        \includegraphics[width=\textwidth]{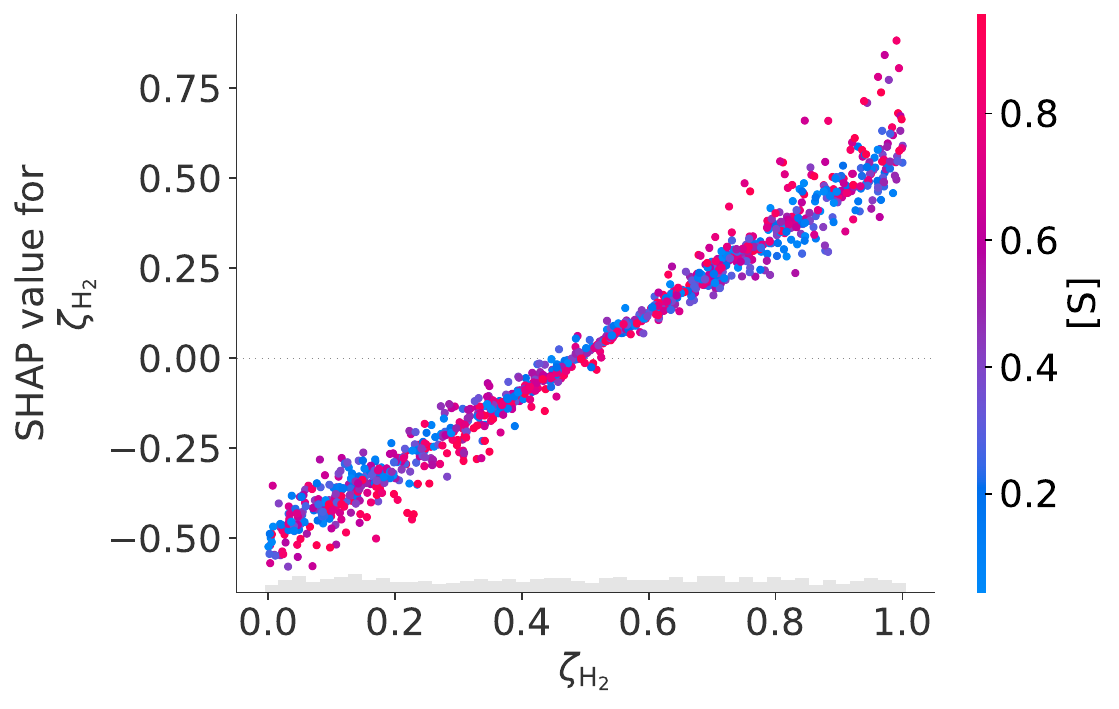}
    \end{subfigure}
    \begin{subfigure}[b]{0.246\textwidth}
        \centering
        \includegraphics[width=\textwidth]{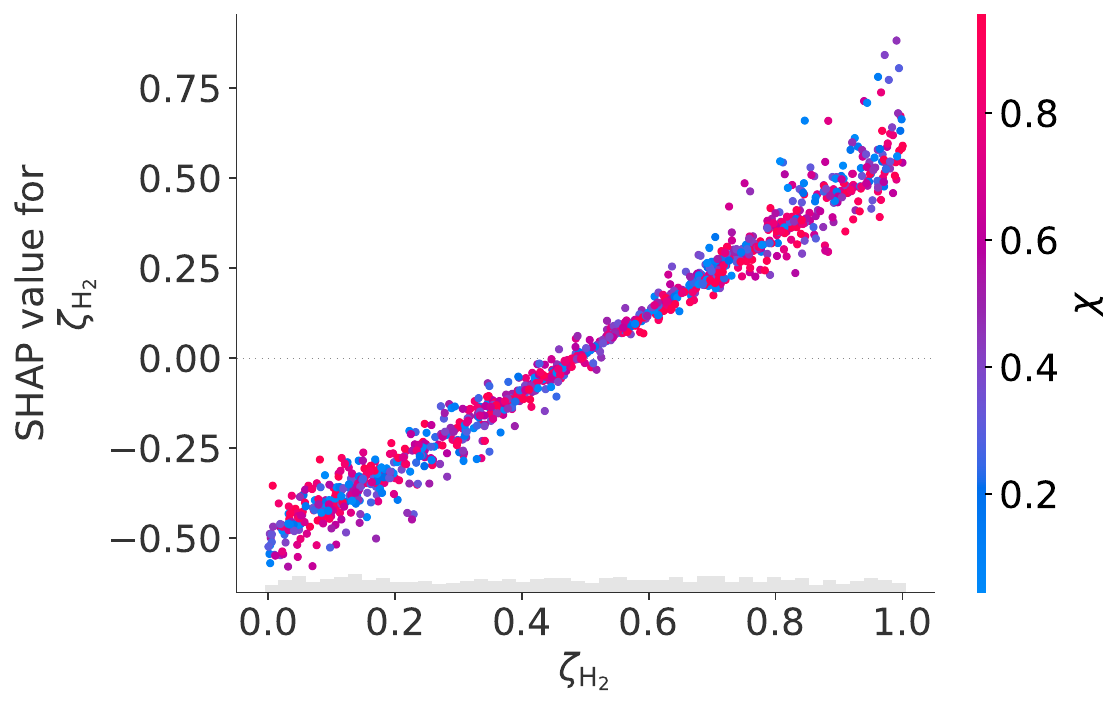}
    \end{subfigure}
    ~
    \begin{subfigure}[b]{0.246\textwidth}
        \centering
        \includegraphics[width=\textwidth]{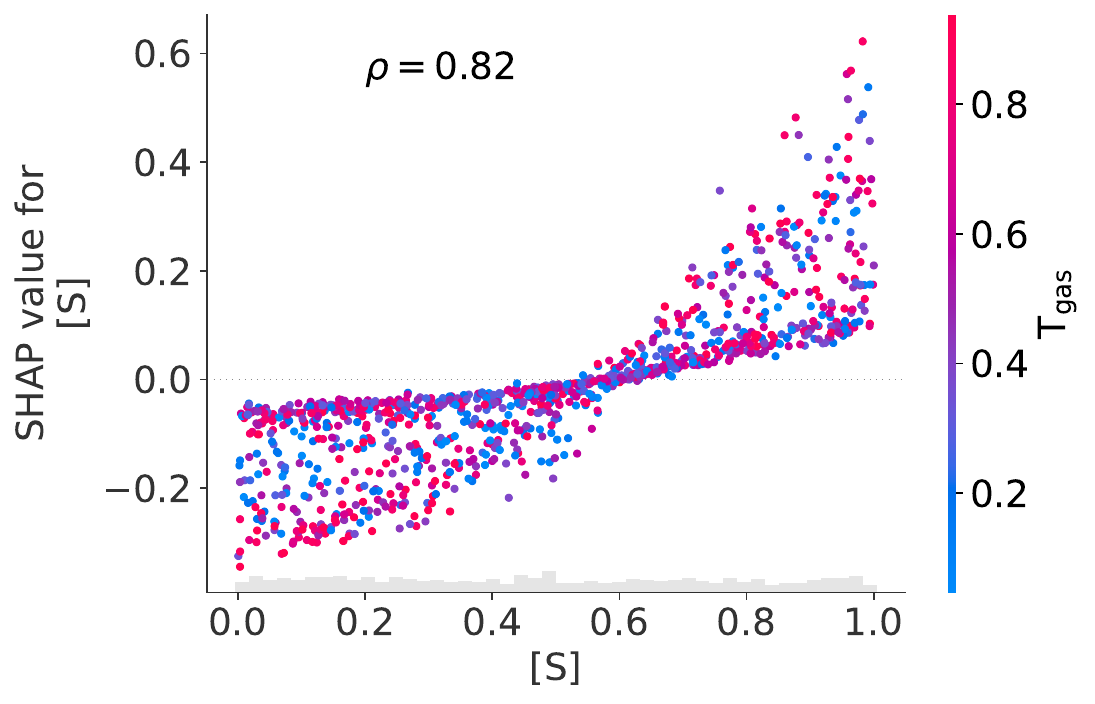}
    \end{subfigure}
    \begin{subfigure}[b]{0.246\textwidth}
        \centering
        \includegraphics[width=\textwidth]{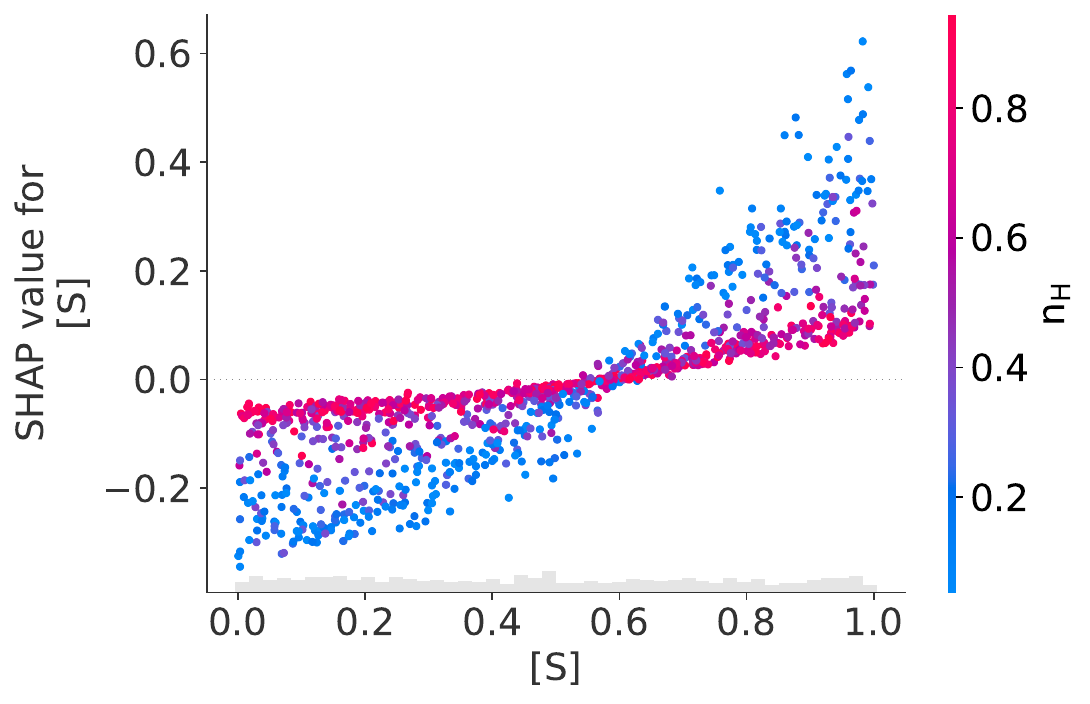}
    \end{subfigure}
    \begin{subfigure}[b]{0.246\textwidth}
        \centering
        \includegraphics[width=\textwidth]{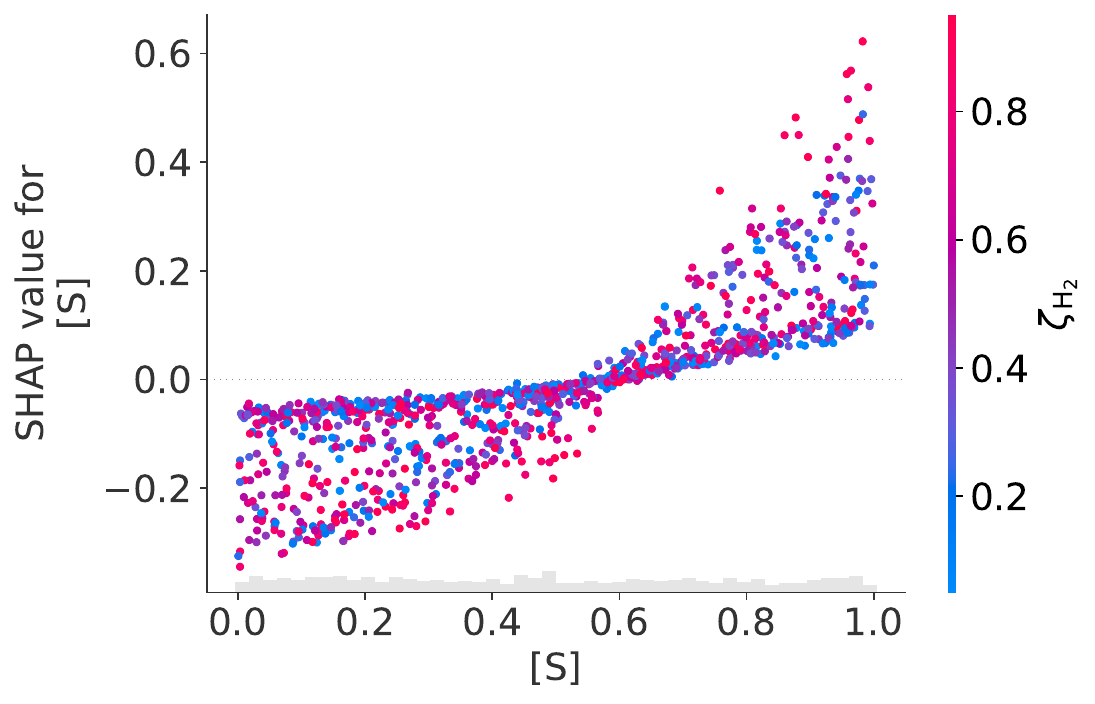}
    \end{subfigure}
    \begin{subfigure}[b]{0.246\textwidth}
        \centering
        \includegraphics[width=\textwidth]{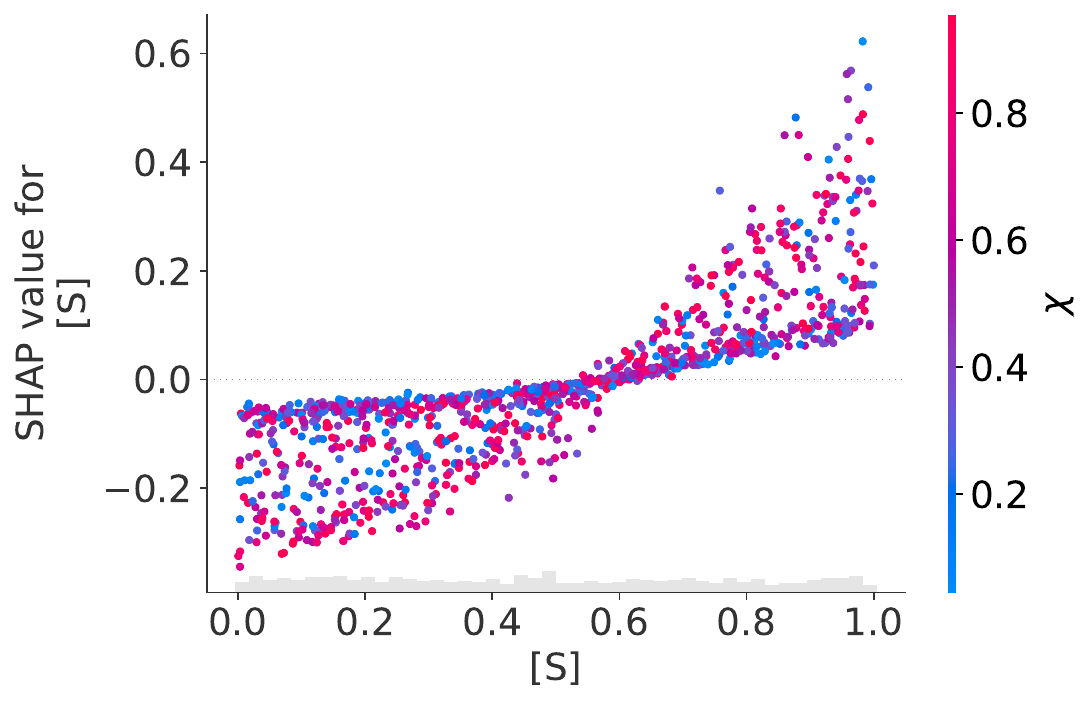}
    \end{subfigure}
    ~
    \begin{subfigure}[b]{0.246\textwidth}
        \centering
        \includegraphics[width=\textwidth]{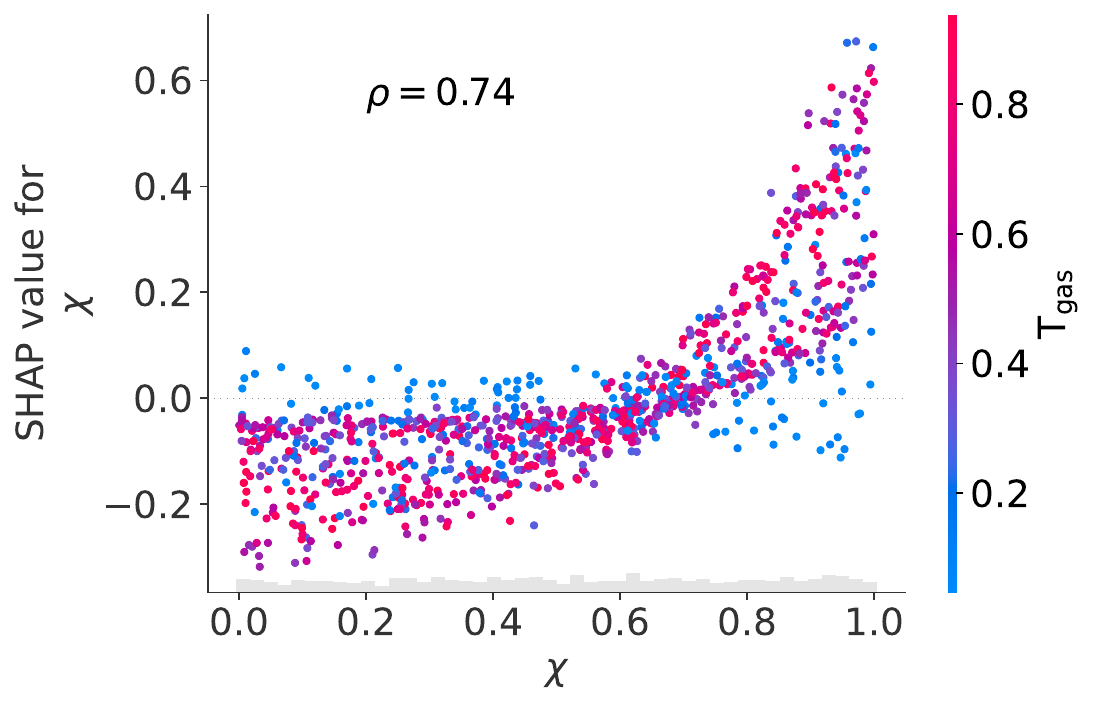}
    \end{subfigure}
    \begin{subfigure}[b]{0.246\textwidth}
        \centering
        \includegraphics[width=\textwidth]{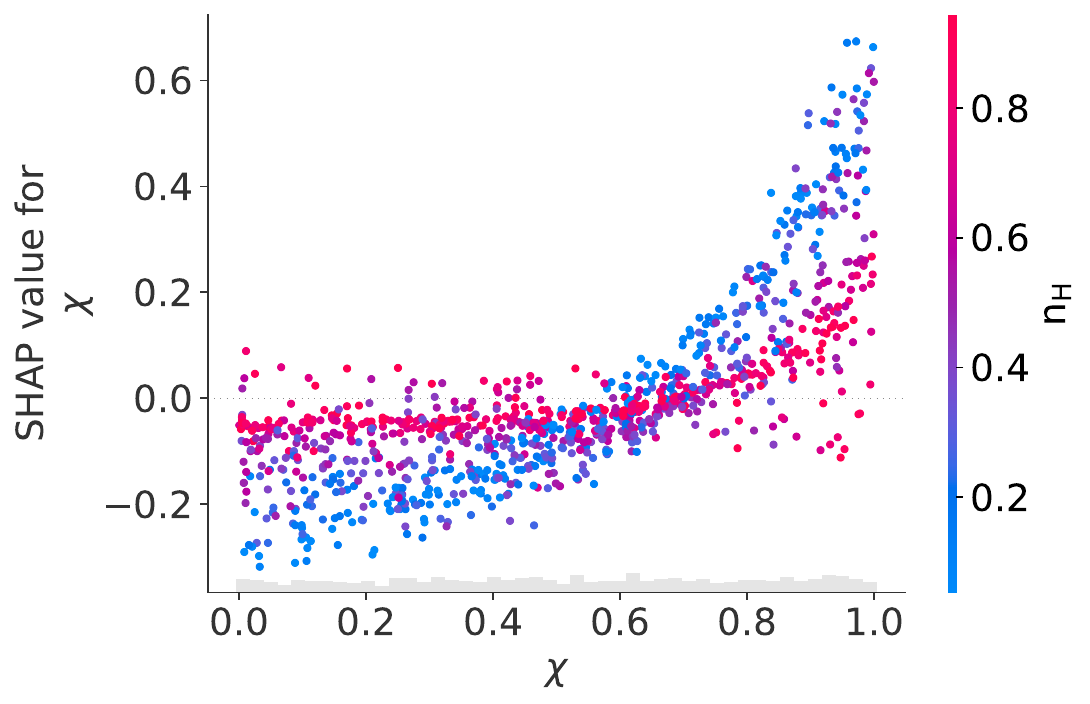}
    \end{subfigure}
    \begin{subfigure}[b]{0.246\textwidth}
        \centering
        \includegraphics[width=\textwidth]{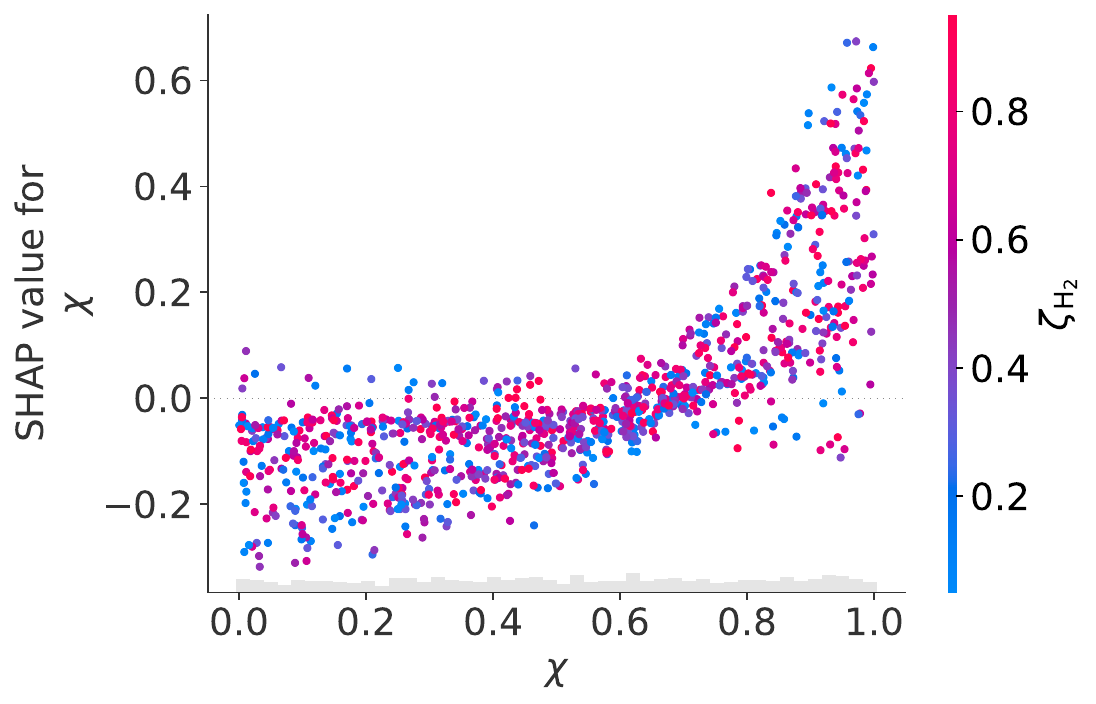}
    \end{subfigure}
    \begin{subfigure}[b]{0.246\textwidth}
        \centering
        \includegraphics[width=\textwidth]{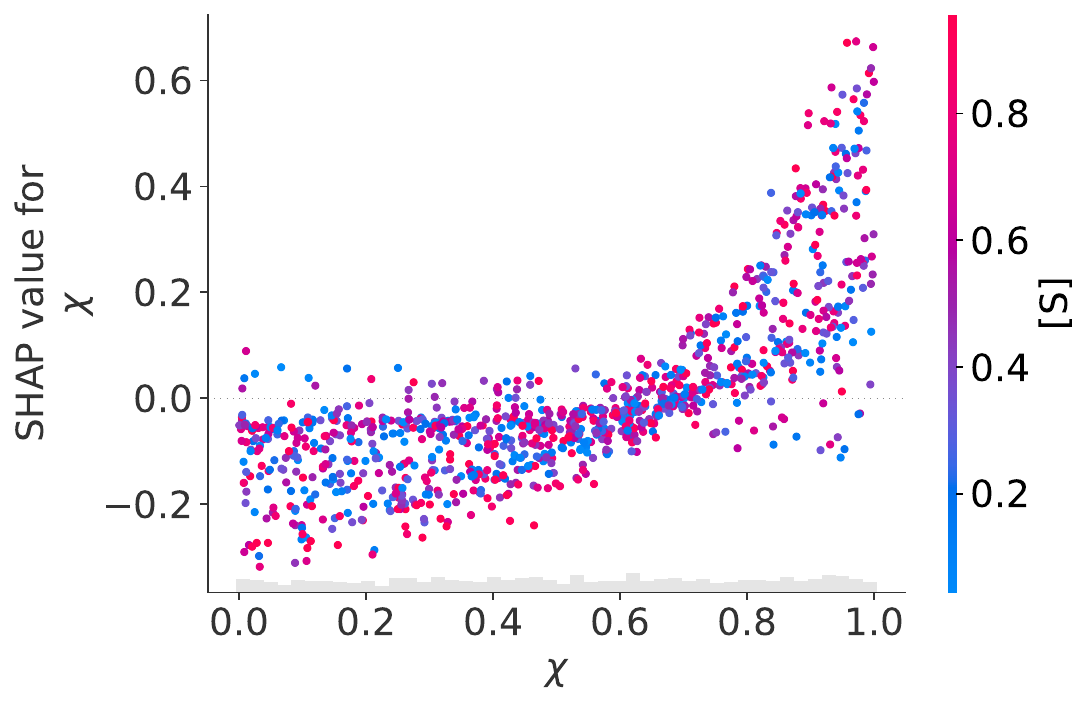}
    \end{subfigure}
    \caption{Correlations between SHAP and feature values of the 2048 predictions of the number of electrons for gas temperature (first row), gas density (second row), cosmic-ray ionization rate (third row), initial sulphur abundance (fourth row), and FUV field strength (fifth row). In each row, the 2048 predictions are color-coded by the rest of the features. The histogram on the x-axis shows the uniform sampling in the range of feature values.}
    \label{fig:correlations}
    \end{figure*}

        	
Sulphur atoms, with an ionization potential of $10.36$ eV, are easily ionized in translucent gas, 
becoming an important donor of electrons in these areas inside molecular clouds 
\citep[see, e.g.,][]{Goicoechea2006, Fuente2019,Bulut2021}. We expect that in denser and 
shielded gas, where sulphur is mostly found in sulphur-bearing molecules, the initial sulphur 
abundance becomes less important to set the ionization degree of the gas. The ionization degree 
is a key quantity that influences the coupling between the gas and the interstellar magnetic 
field. Consequently, it strongly impacts the dynamics of the gas and, potentially, the star 
formation process. With the advent of numerical simulations of core collapse, this quantity is 
present in diffusive phenomena like ambipolar diffusion, Ohmic dissipation, and Hall effect 
\citep{Masson2012}. These diffusive phenomena were proposed as a solution to the magnetic 
braking catastrophe, in which the implementation of ideal magnetohydrodynamics (MHD) prevents 
disks from being formed \citep[see, e.g.,][]{Allen2003, Galli2006, Hennebelle2008, Commercon2010}. 
It is therefore interesting to examine the impact of the initial sulphur abundance on the number 
of electrons for physical conditions corresponding to different layers of a molecular cloud and 
discuss the possible implications in the star formation process.
        	
To investigate the impact of the model features on the number of electrons, we used 
the Python package SHAP \citep{Lundberg2017}. We first created a wrapper around the 
emulator so that the input features are scaled to the range $[0,1]$ and the output is the 
decimal logarithm of the chemical abundances at a given time instant. Unless specified, the 
values of the output are taken at 1 Myr. We performed a random uniform sampling of the parameter 
space with 2048 predictions. The SHAP values of the five features for the 2048 predictions and 
192 chemical species were calculated with the Explainer object of the SHAP library. The importances 
of the features (see Eq. \ref{eq:importance}) setting the number of electrons are shown in 
Fig. \ref{fig:rankBeeSwarm}. According to the ranking of features (left hand side of Fig. \ref{fig:rankBeeSwarm}), 
the density is the feature that makes the greatest impact on the number of electrons, with a contribution 
of $51.2\%$ (Eq. \ref{eq:relImportance}), followed by the cosmic ray ionization rate ($21.6\%$), 
the FUV field strength ($9.6\%$), and the initial sulphur abundance and gas temperature, both with 
a contribution of $8.8\%$. In the right hand side of Fig. \ref{fig:rankBeeSwarm} we show the 
beeswarm plot of the ionization degree, that is, the SHAP values for all the features that correspond 
to each one of the 2048 randomly-sampled predictions, color-coded by the feature value. Our results 
indicate that SHAP values for density are anti-correlated with the feature (density) value, that is, 
high gas densities have a negative impact on the number of electrons. This is in agreement with our 
previous discussion: as density increases, sulphur is no longer in its ionized form but found in 
sulphur bearing molecules, thus reducing its contribution to the ionization state of the gas. For the 
rest of features, this behavior is inverted, as their SHAP values are correlated with their value. The 
cosmic-ray ionization rate and the FUV field strength have a positive correlation with their 
corresponding SHAP values. As expected, stronger ionizing agents have a positive impact in the ionization 
of the gas. Likewise, higher temperatures are associated with a positive impact on the number of 
electrons of the gas, although this correlation is weaker.

Correlations are quantified in Fig. \ref{fig:correlations}. A strong correlation, $\rho=0.99$, 
is seen between the SHAP and feature values of the cosmic-ray ionization rate, while a strong 
anti-correlation, $\rho=-0.97$, is present between the SHAP and feature values of the gas density. 
Correlations on the rest of features are weaker, meaning that predictions with a similar feature 
value may have different SHAP values for that feature. This indicates that the SHAP value for a 
feature may not be only dependent on that feature but on several interacting features. We investigate 
feature interaction in Fig. \ref{fig:correlations}, showing the pairs of SHAP and feature values for 
a given feature, color-coded by the rest of features. Gas density $n_{\rm H}$ shows a clear 
interaction with the initial sulphur abundance $[{\rm S}]$ and the FUV field strength $\chi$ (see 
plots in the second column, fourth and fifth row, respectively, in Fig. \ref{fig:correlations}). In 
the case of the initial sulphur abundance, for any given value of this feature, lower densities yield 
higher (in absolute value) SHAP values for $[{\rm S}]$ than higher gas densities. Since SHAP values 
tell us how each feature contributes to the deviation of a given prediction from the average outcome 
(Eq. \ref{eq:additiveExplanation}), this indicates that the number of electrons is susceptible to 
change more when the initial sulphur abundance is modified in low-density gas than when it is modified 
in denser gas, where the number of electrons is more robust against changes in $[{\rm S}]$. A similar 
behavior is also observed in the interaction between density $n_{\rm H}$ and the FUV field strength $\chi$. 

We can provide a quantitative estimation of how different densities alter the variations in the ionization 
degree when the initial sulphur abundance is modified. We first considered the electron
density $n_e$ in two low-density scenarios, with low and high initial
sulphur abundances respectively. The family of vectors of input
parameters with low density and low initial sulphur abundance,
both normalized to the range $[0,1]$, following the standard ordering of features
given by the set $N=\{T,n_\mathrm{H},\zeta_{\mathrm{H}_2},\left[S\right],\chi\}$, is given by
\begin{equation}
\vect \theta^{a} =[T, 0, \zeta_{\mathrm{H}_2},0,\chi ]
\label{eq:low}
\end{equation}
with temperature $T$, cosmic-ray ionization rate $\zeta_{\mathrm{H}_2}$, and FUV field
strength $\chi$ set as free parameters. Similarly, the family of vectors
of input parameters with low density and high initial sulphur
abundance is given by
\begin{equation}
\vect \theta^{a} =[T, 0, \zeta_{\mathrm{H}_2},1,\chi ].
\label{eq:high}
\end{equation}
We estimated the electron density in each case by averaging
over the free parameters using a set of 30000 uniformly sampled
vectors of input parameters following Eqs. \ref{eq:low} and \ref{eq:high}. The resulting
electron densities are:
\begin{align*}
    \begin{split}
    & \langle \log{n_{e}}(\vect \theta^{a}) \rangle = -6.53,\\
    & \langle \log{n_{e}}(\vect \theta^{b}) \rangle = -5.50,
    \end{split}
\end{align*}
yielding the following relative difference in the electron density for the low density case
\begin{equation}\label{eq:relDiffLowDens}
\frac{\langle n_{e}(\vect \theta^{b}) \rangle}{\langle n_{e}(\vect \theta^{a}) \rangle } = 10^{-5.50+6.53}\sim 10.
\end{equation}
That is, in the low density scenario ($n_{\rm H} \sim 10^{4}$ cm$^{-3}$), increasing the 
initial sulphur abundance from $[S] = 8\times 10^{-8}$ to $[S] = 1.5\times 10^{-5}$ leads 
to an enhancement of the electron density by an order or magnitude. We can perform a 
similar analysis in a high density scenario ($n_{\rm H} = 10^{7}$ cm$^{-3}$, normalized 
density of 1). Given the following two families of input vectors:
\begin{equation*}\label{eq:vecHighDensLowS}
    \vect \theta^{c} =\ \left[T, 1, \zeta_{\mathrm{H}_2}, 0, \chi\right],
\end{equation*}
and
\begin{equation*}\label{eq:vecHighDensHighS}
    \vect \theta^{d} =\ \left[T, 1, \zeta_{\mathrm{H}_2}, 1, \chi\right],
\end{equation*}
with temperature $T$, cosmic-ray ionization rate $\zeta_{\mathrm{H}_2}$, and FUV field 
strength $\chi$ set as free parameters and averaged over 30000 random samples, their respective electron densities are:
\begin{align*}
    \begin{split}
    & \langle \log{n_{e}}(\vect \theta^{c}) \rangle = -8.60,\\
    & \langle \log{n_{e}}(\vect \theta^{d}) \rangle = -8.57.
    \end{split}
\end{align*}
This results in a relative difference of
\begin{equation*}
    \frac{\langle n_{e}(\vect \theta^{d}) \rangle}{\langle n_{e}(\vect \theta^{c}) \rangle} = 10^{-8.57+8.60}\sim 1.07,
\end{equation*}
that is, in the high density scenario, the density of electrons is only increased 
by a 7\% when the initial sulphur abundance is enhanced from $[S] = 7.5\times 10^{-8}$ to $[S] = 1.5\times 10^{-5}$.

\begin{figure}
    \centering
    \includegraphics[width=0.48\textwidth]{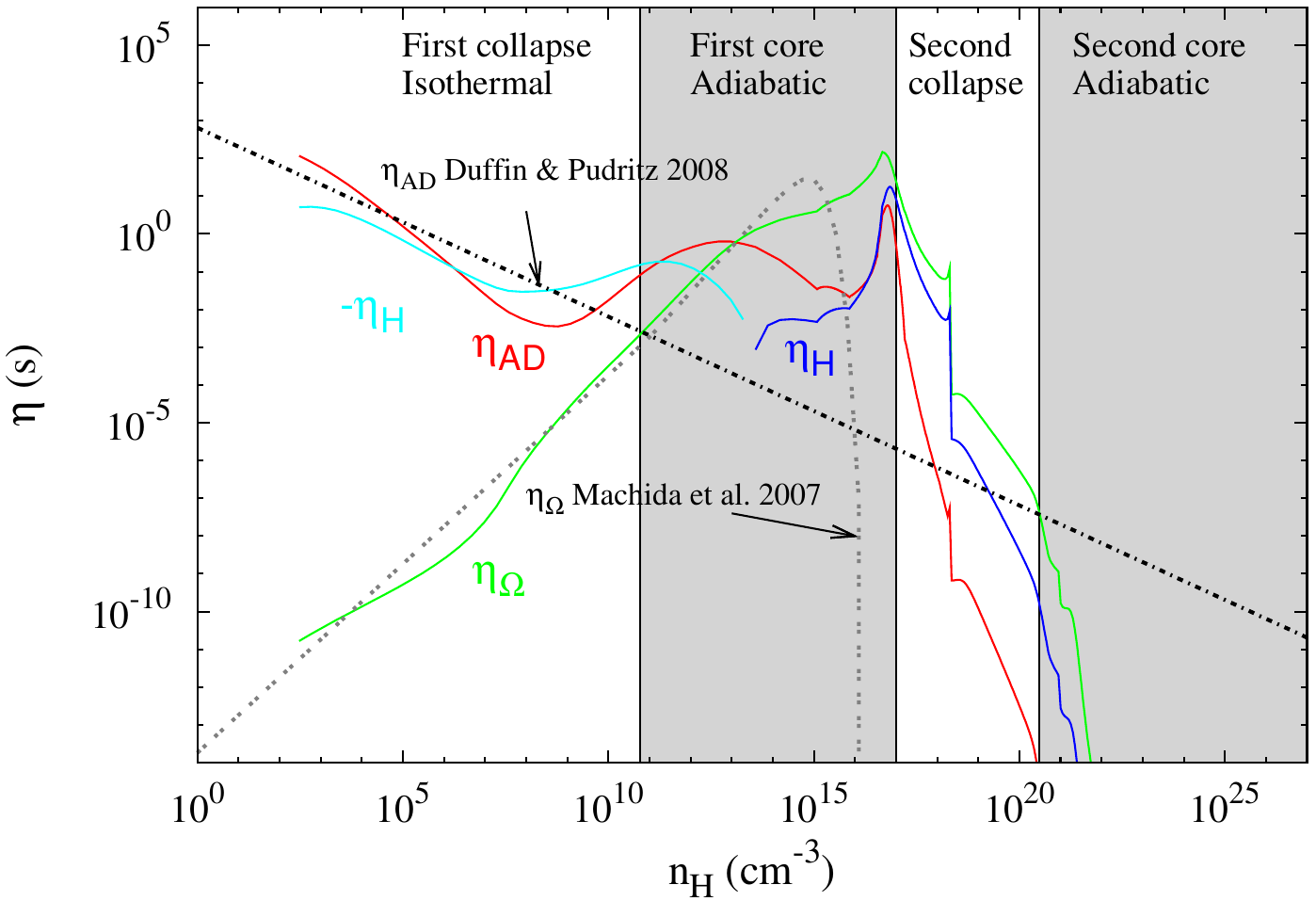}
    \caption{Magnetic resistivities and their relevance at different stages of the star formation 
    process as a function of density. Light blue: negative Hall resistivity, dark blue: positive Hall 
    resistivity, red: ambipolar diffusion resistivity, and green: Ohm resistivity. Taken from \citet{Marchand2016}.}
    \label{fig:resistivities}
\end{figure}
        		
It is now worth discussing whether the link between sulphur abundance, density, and the ionization 
degree has relevant implications in the star formation process. As mentioned above, non-ideal MHD 
simulations are the state-of-the-art tool to model core collapse and star formation. They include 
diffusive phenomena like ambipolar diffusion, Ohmic dissipation, and Hall effect that regulate the 
dissipation of the magnetic flux \citep{Masson2012}. These resistivities depend on the ionization 
degree of gas and dust \citep[see, e.g.,][]{Marchand2016}. At densities $n_{\rm H} < 10^7$ cm$^{-3}$, 
electrons are the dominant charge carriers while at higher densities metallic ions and grains overcome 
electrons becoming the main charged species \citep{Marchand2016}. This density is precisely the largest 
considered in our parameter space and, accordingly, the possible effects of the initial sulphur 
abundance on the number of electrons and the dynamics of star formation should start early on in 
the isothermal core collapse. In this stage, in the low density scenario $n_{\rm H} \sim 10^4$ cm$^{-3}$, 
the initial sulphur abundance can be more effective increasing or decreasing the number of electrons 
and, consequently, modifying the magnetic resistivities. This is particularly relevant for the role of 
ambipolar diffusion in the star formation process (see Fig. \ref{fig:resistivities}): a high initial 
sulphur abundance in a low-density medium leads to a higher ionization of the gas that becomes better 
collisionally coupled with the neutral species, reducing the ambipolar diffusion resistivity. Likewise, 
low initial sulphur abundances could lead to a lower electron abundance that enhances ambipolar 
resistivity. As density increases, the initial sulphur abundance loses efficiency at setting the amount 
of electrons, and eventually, metallic ions and grains become the main charge carriers. We conclude that, 
at the early times during the isothermal collapse that precedes the First Hydrostatic Core phase, the 
initial sulphur abundance can modify the ambipolar resistivity effectively in low density environments. 
This ability is greatly reduced as collapse proceeds and the amount of electrons decreases. While the 
ambipolar diffusion does not seem to have an impact on the mass and radius of the first Larson core when 
compared to the ideal MHD case \citep{Masson2016}, it defines the ambipolar-diffusion timescale, the 
timescale in which filaments fragment and cores form \citep{Ciolek1993, Tassis2004, Kudoh2014, Burge2016}. In the simulations carried out by \citet{Burge2016}, the collapse rate is inversely proportional to the ionization degree. According to their results, a difference of an order of magnitude in the ionization degree, which is the difference we predict between the high and low initial sulphur abundance scenarios for the low 
density case (see Eq. \ref{eq:relDiffLowDens}), leads to different collapse dynamics: a magnetically-regulated collapse, where gas is strongly coupled to the magnetic field and the gas collapses at velocities much lower than the sound speed, and a gravity-dominated one where gas is weakly coupled to the magnetic field and reaches higher velocities. Consequently, the initial sulphur abundance and its impact on the ionization state of the gas in low density environments could potentially play a role in the dynamics of core collapse.

\section{Conclusions}
Astrochemistry is a fundamental tool for the interpretation of observations in the interstellar medium.
The solution to the system of non-linear coupled ordinary differential
equations describing the time-dependent chemistry in large chemical networks 
is feasible nowadays but requires lots of computing power. Having access to a fast
emulator will open the possibility of improving our understanding of the 
physical conditions in the interstellar medium. 

We have developed such an emulator making use of NFs conditioned on the physical
conditions and trained with computations carried out with the Nautilus code. We 
have demonstrated, through validation studies, that the emulator
does an excellent job in approximating the time evolution of 192 species. The logarithm
of the abundance of almost all species can be obtained with uncertainties well below 0.2 dex
for all times below 10$^7$ yr. The resulting model is able to provide the time evolution
for all species $\sim 10^4$ times faster than Nautilus. In addition, the computing time is not sensitive to the specific
physical conditions, carrying out exactly the same number of operations for any combination
of the input parameters inside the ranges defined in Tab. \ref{tab:phys}. 
This allows us to perform thousands of simulations in
a reasonable computing time to fit spectroscopic observations or explore
the influence of different input parameters in chemical predictions. In this paper, we have
used Shapley values to investigate the influence of the sulphur elemental abundance on the gas ionization degree,
key parameter in the dynamical evolution of interstellar clouds.
We conclude that the initial sulphur abundance can modify the ambipolar difussion resistivity in low 
density environments (n$_{\rm H}$$\sim$10$^4$  cm$^{-3}$ to a few 10$^5$ cm$^{-3}$), hence
determining the timescale for fragmentation and collapse. 
This ability is reduced as collapse proceeds and density achieves values 
close to n$_{\rm H}$$\sim$10$^7$ cm$^{-3}$. In this high density regime the influence of sulphur abundance
on the dynamics is negligible.

Additionally, the model is fully differentiable. This opens the possibility of including the chemistry in
more elaborate forward models, to better understand the observations. One example
is the Bayesian inference of physical properties in the interstellar medium 
via the interpretation of observed spectral lines. The speed of our model make it feasible
to systematically determine the physical and chemical conditions over large regions of the sky
to characterize star forming regions of ours and external galaxies.

Emulators, obviously, trade speed for flexibility. Since these models require a training set, the chemical
network, the number of species, the range of physical parameters and all remaining
hyperparameters have to be fixed in advance. Updating the chemical network (either changing
the reaction rates or adding/removing reactions), including
more species or exploring different regions of the space of parameters requires
a retraining of the neural model. Anyway, the training time is negligible
when compared with the amount of time saved in any downstream task carried out
with the emulator.

\section*{Acknowledgements}
AAR thanks Carlos D\'{\i}az Baso for helpful discussions on the training of NFs.
AAR acknowledges support from the Agencia Estatal de Investigaci\'on del Ministerio de Ciencia, Innovaci\'on
y Universidades (MCIU/AEI) and the European Regional Development Fund (ERDF) through project PID2022-136563NB-I00.
DNA acknowledges funding support from Fundaci\'on Ram\'on Areces through their 
international post-doc grant program. AF and PRM thank Spanish MICIN for funding 
through project PID2019-106235GB-I00 and grant PID2022-137980NB-I00 by the Spanish 
Ministry of Science and Innovation/State Agency of Research MCIN/AEI/ 10.13039/501100011033 
and by “ERDF A way of making Europe”. AF thanks the European Research Council (ERC) for 
funding under the Advanced Grant project SUL4LIFE, grant agreement No101096293.
This research made use of the IAC HTCondor facility (http://research.cs.wisc.edu/htcondor/), partly 
financed by the Ministry of Economy and Competitiveness with FEDER funds, code IACA13-3E-2493.
This research has made use of NASA's Astrophysics Data System Bibliographic Services.
We acknowledge the community effort devoted to the development of the following 
open-source packages that were
used in this work: \texttt{numpy} \citep[\texttt{numpy.org},][]{numpy20}, 
\texttt{matplotlib} \citep[\texttt{matplotlib.org},][]{matplotlib}, \texttt{PyTorch} 
\citep[\texttt{pytorch.org},][]{pytorch19}, \texttt{zarr} (\texttt{github.com/zarr-developers/zarr-python})
and \texttt{SHAP} \citep[\texttt{github.com/shap/shap}]{Lundberg2017}.

\section*{Data Availability}
Access to the data and codes used in this work is described in the repository \texttt{https://github.com/aasensio/neural\_chemistry}.



\bibliographystyle{mnras}




\appendix

\section{Quantitatve results for all molecules}
\begin{figure*}
    \centering
    \includegraphics[width=\textwidth]{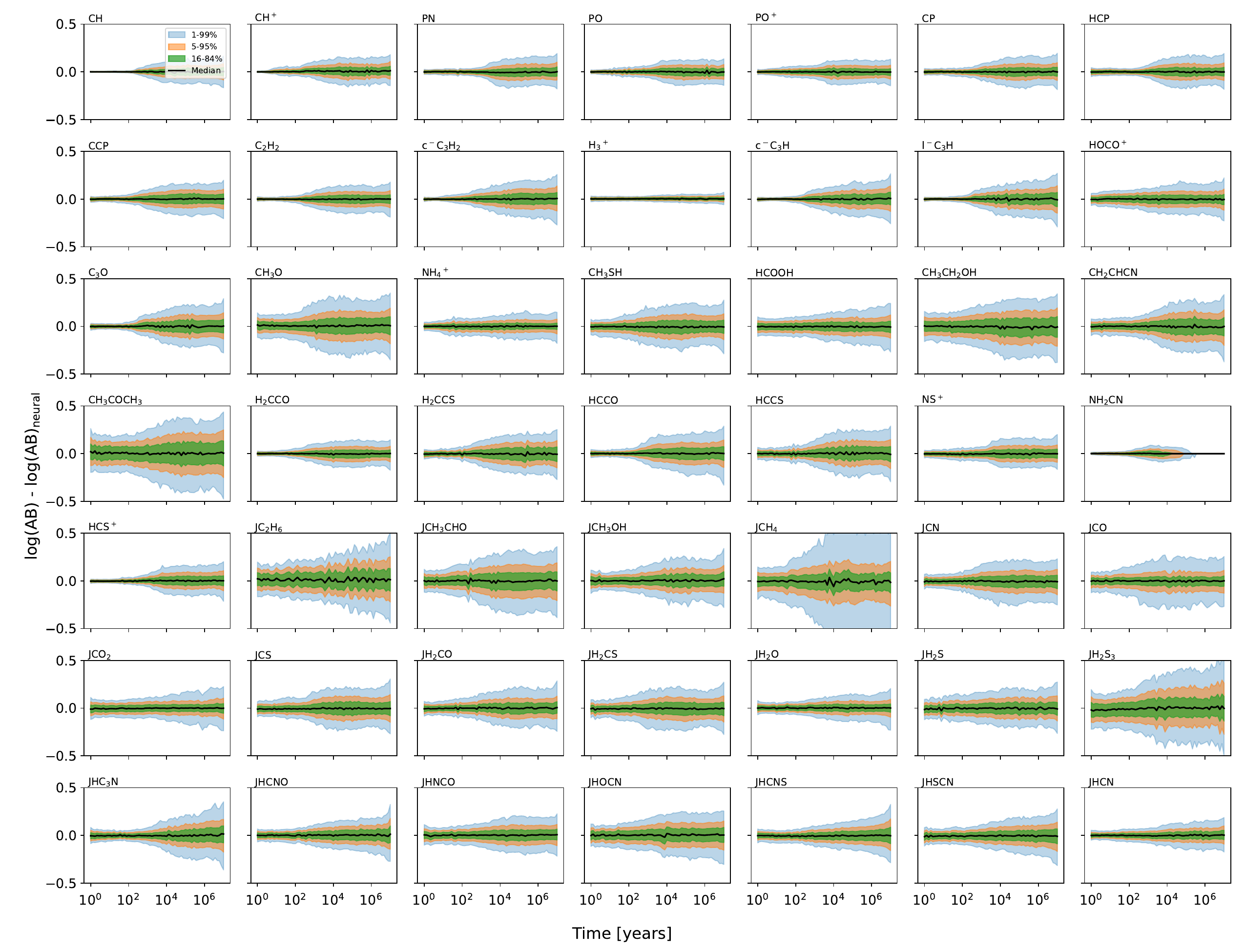}
    \caption{Difference in logarithm between the emulator and the original run
    for selected species. The solid line shows the median difference, while 
    the shaded area shows the quantiles 1-99 in blue (approximately $3\sigma$), 
    5-95 (approximately $2\sigma$) in orange and 16-84 (approximately $1\sigma$)
    in green.
     \label{fig:stats1}}
    \end{figure*}

\begin{figure*}
    \centering
    \includegraphics[width=\textwidth]{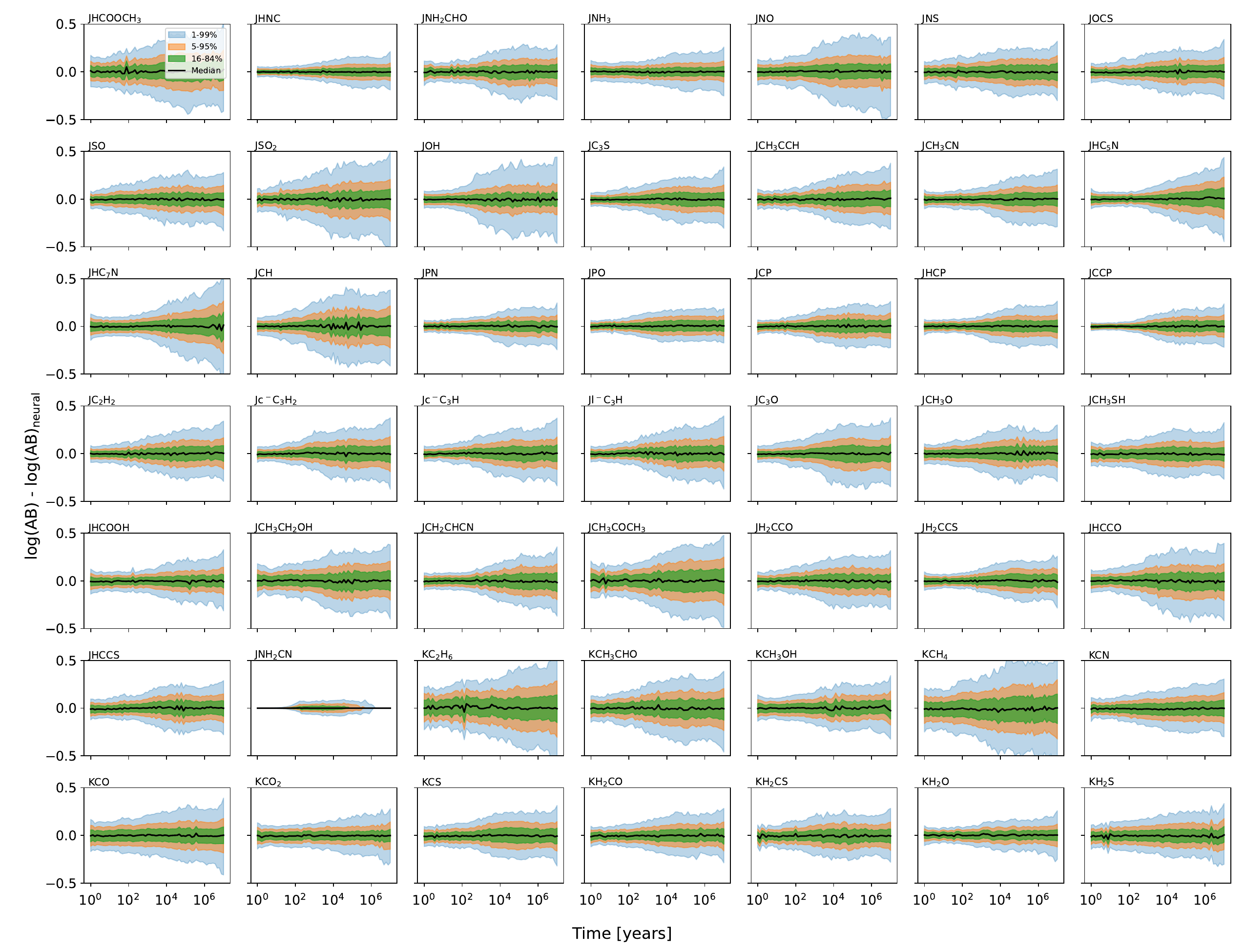}
    \caption{Difference in logarithm between the emulator and the original run
    for selected species. The solid line shows the median difference, while 
    the shaded area shows the quantiles 1-99 in blue (approximately $3\sigma$), 
    5-95 (approximately $2\sigma$) in orange and 16-84 (approximately $1\sigma$)
    in green.
        \label{fig:stats2}}
    \end{figure*}

\begin{figure*}
    \centering
    \includegraphics[width=\textwidth]{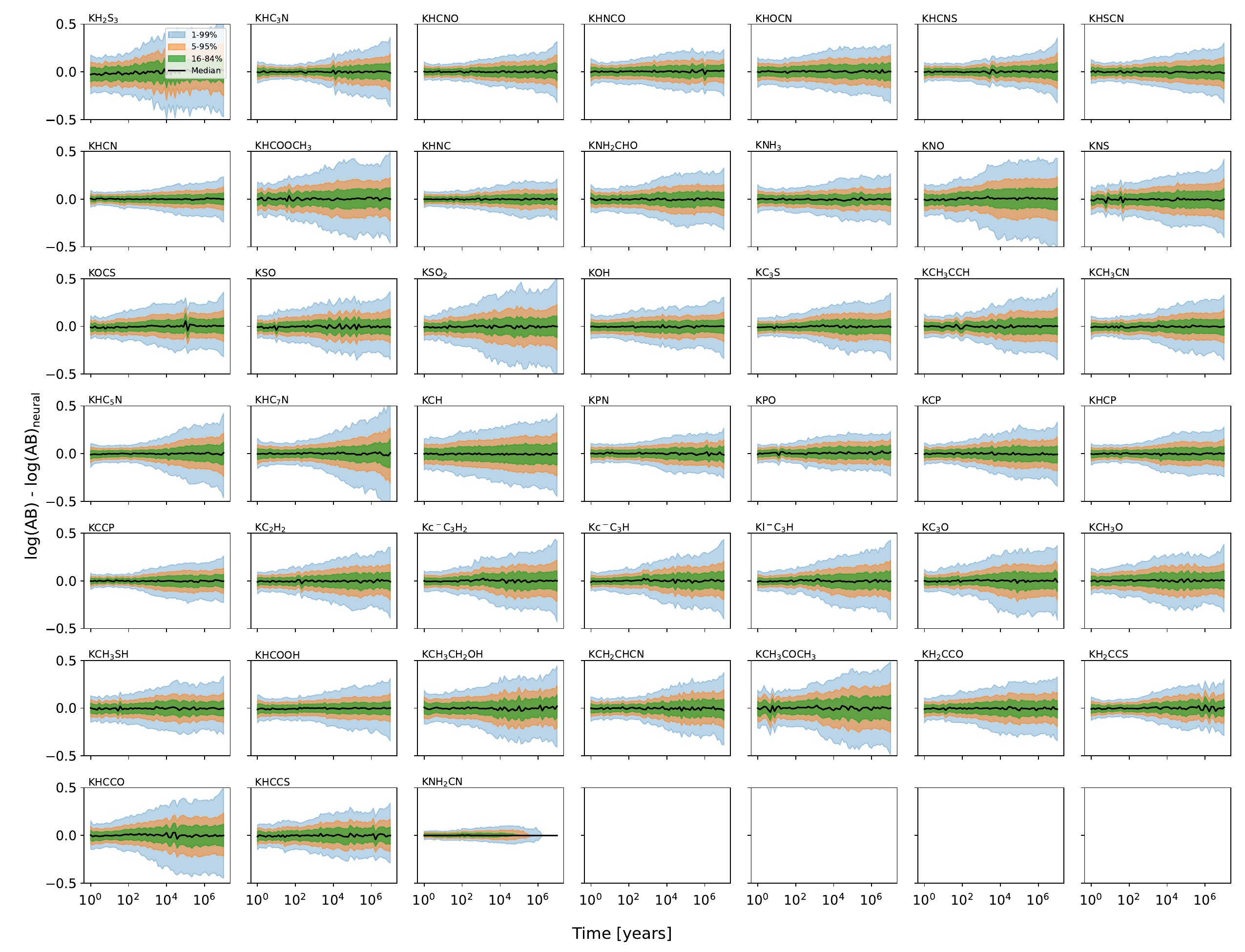}
    \caption{Difference in logarithm between the emulator and the original run
    for selected species. The solid line shows the median difference, while 
    the shaded area shows the quantiles 1-99 in blue (approximately $3\sigma$), 
    5-95 (approximately $2\sigma$) in orange and 16-84 (approximately $1\sigma$)
    in green.
        \label{fig:stats3}}
    \end{figure*}

Figs. \ref{fig:stats1}, \ref{fig:stats2} and \ref{fig:stats3} display the distribution of differences in 
logarithm for all species in the validation set, which complement those shown in Fig. \ref{fig:stats}.

\bsp	
\label{lastpage}
\end{document}